\documentclass[journal]{IEEEtran}

\usepackage{cite}

\ifCLASSINFOpdf
\else
\fi
\usepackage{amsmath}
\ifCLASSOPTIONcompsoc
  \usepackage[caption=false,font=normalsize,labelfont=sf,textfont=sf]{subfig}
\else
  \usepackage[caption=false,font=footnotesize]{subfig}
\fi
\usepackage{booktabs}
\usepackage{fancyvrb}
\usepackage{enumitem}
\usepackage{array}
\usepackage{color}
\usepackage[flushleft]{threeparttable}
\usepackage{siunitx}
\usepackage{graphicx}
\usepackage{amsmath}
\usepackage{mathtools}
\usepackage{amssymb}
\usepackage{bbm}
\usepackage{comment}
\usepackage{multirow} 
\usepackage{color}
\usepackage[normalem]{ulem}
\usepackage[table]{xcolor}
\usepackage{pgfplots}

\definecolor{darkbrown}{rgb}{0, 0, 0.7}
\newcommand{\resizenet}{\textcolor{black}{ResizeParamNet}}
\newcommand{\rc}{\textcolor{black}{Resize-Compress}}
\newcommand{\pZ}{\phantom{0-}}

\newcommand\norm[1]{\left\lVert#1\right\rVert}

\definecolor{Vermillion}{RGB}{213, 94, 0}

\begin{document}

\title{Estimating the Resize Parameter in End-to-end Learned Image Compression}

\author{Li-Heng~Chen,
        Christos~G.~Bampis,
        Zhi~Li,
        Luk\'a\v s~Krasula,
        and~Alan~C.~Bovik,~\IEEEmembership{Fellow,~IEEE}
\thanks{L.-H.~Chen and A.~C.~Bovik are with the Department of Electrical and Computer Engineering, University of Texas at Austin, Austin, TX, 78712 USA (email:lhchen@utexas.edu; bovik@ece.utexas.edu).}
\thanks{C.~G.~Bampis, Z.~Li, and L.~Krasula are with Netflix Inc. Los Gatos, CA, 95032 USA (email:christosb@netflix.com; zli@netflix.com; lkrasula@netflix.com).}
\thanks{This work is supported by Netflix and by grant number 2019844 for the National Science Foundation AI Institute for Foundations of Machine Learning (IFML).}
}
\markboth{}%
{Shell \MakeLowercase{\textit{et al.}}: Bare Demo of IEEEtran.cls for IEEE Journals}
\maketitle

\begin{abstract}
We describe a search-free resizing framework that can further improve the rate-distortion tradeoff of recent learned image compression models. Our approach is simple: compose a pair of differentiable downsampling/upsampling layers that sandwich a neural compression model. To determine resize factors for different inputs, we utilize another neural network jointly trained with the compression model, with the end goal of minimizing the rate-distortion objective. Our results suggest that ``compression friendly'' downsampled representations can be quickly determined during encoding by using an auxiliary network and differentiable image warping. By conducting extensive experimental tests on existing deep image compression models, we show results that our new resizing parameter estimation framework can provide Bj\o{}ntegaard-Delta rate (BD-rate) improvement of about 10\% against leading perceptual quality engines. We also carried out a subjective quality study, the results of which show that our new approach yields favorable compressed images. To facilitate reproducible research in this direction, the implementation used in this paper is being made freely available online at: https://github.com/treammm/ResizeCompression.
\end{abstract}


\IEEEpeerreviewmaketitle

\section{Introduction}
\IEEEPARstart{W}{hen} properly applied, spatial resolution reduction is a simple and natural way to compress visual information. It plays a significant role in many multimedia compression standards and applications. For example, it is common to decimate the chroma components of an image or video to reduce bitrate consumption before encoding a video, then upsize them before display. Many studies \cite{winkler2001book,Uhrina2017} have concluded that this fixed kind of sub-sampling operation results in little impact on perceived quality, because of the smaller bandwidth of chromatic information, and human perception of it \cite{Mullen85}. In recent streaming video workflows, non-normative adaptive spatial resolution changes of both luma and chroma are now widely used by service providers like Netflix and Youtube to improve bandwidth consumption while maintaining the quality of experience of viewers \cite{cvx_pertitle15, CChen2018, Wu2020}. Moreover, recent advanced video coding standards, such as AOMedia Video 1 (AV1) \cite{ChenAV12020} and Versatile Video Coding (VVC) \cite{BrossVVC2021}, have adopted the concept of spatial resizing as an in-loop coding tool \cite{JoshiAV1SR2019, JVETM0135}. Despite its usefulness, an important barrier to optimizing video resizing in compression workflows is the lack of an automatic protocol for finding the best reduced resolution in the perceptual rate-distortion sense. Current methods rely on exhaustive search over a set of resolutions.

\begin{figure}[!t]
  \centering
  \footnotesize
  \includegraphics[width=3.49in]{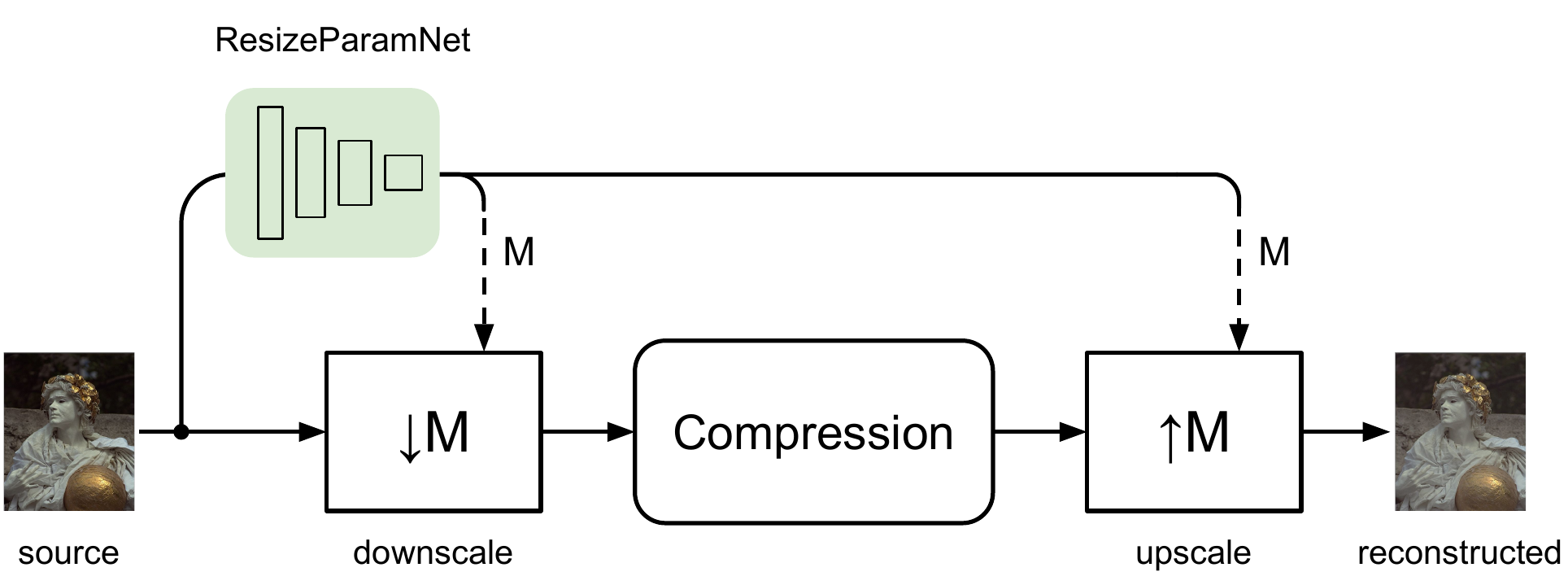}
  \caption{Proposed framework for enhancing \textit{learned} image compression using adaptive rescaling: additional downsampling ($\downarrow$M) and upsampling ($\uparrow$M) blocks are added before and after the compression network, respectively. The parameter $\mathit{M}\in(0,1]$ is the scale factor, which is estimated by an auxiliary network trained in an end-to-end manner. Note that $\mathit{M}$ is also signaled in the encoded bitstream.}
  \label{fig:idea}
\end{figure}

In parallel with the recent successes of deep learning on many video processing problems \cite{BurgerSH12, liu2019cyclicgen, SPaul2020, SPaul2021}, a number of promising learning-based image and video compression models have been realized over the past few years. Unlike traditional hybrid codecs like H.264/AVC, which heavily rely on pipelines of hand-designed modules, most learned compression models deployed deep autoencoders that are optimized end-to-end on large datasets. A significant amount of research has been directed towards growing compression model capability and capacity, by modifying network architecture or by adopting detailed entropy models. The idea of incorporating directed spatial resolution changes into learned compression models has not received much attention, despite successes attained on conventional video coding.

Our aim here is to explore the potential of leveraging spatial resizing in learned lossy codecs. To overcome the aforementioned cost of computationally inefficient searches over the space of resize factors, we propose a framework that simultaneously learns the compression model and an estimate of the resize factor. As depicted in Fig. \ref{fig:idea}, the main idea is to regress on the resize factor $M$ using an auxiliary neural network, which we will refer to as \resizenet. Since the source image is the input to the auxiliary network, the resize factor $M$ is estimated in a content-adaptive fashion, without the need for time-consuming search procedures. The input source image is downscaled before compression, then upscaled during reconstruction. Both resizing modules are controlled by the factor $M$. Our approach enables better rate-distortion performance without modifying the compression kernel. We summarize the key characteristics and contribution of this work as follows:
\begin{itemize}
  \item \textit{Generalizability:} We show a way to inject optimized capability as a ``booster'' for learned image compression. This model, which we call \rc, can be used to jointly optimize differentiable resizing layers and a content-dependent resize factor estimator, and can be used with any compression models.
  \item \textit{Efficiency:} Instead of using brute-force search, an auxiliary lightweight CNN (\resizenet) is employed at the encoder side to quickly determine an optimal resize factor for a given input image. The estimated resize parameter is then signaled to the bitstream, with minimal bit overhead.
  \item \textit{(Perceptual) Efficacy:} We comprehensively validate our proposed approach on a variety of representative deep image compression models, demonstrating that it leads to significant improvements in RD performance as measured by a variety of quality metrics. We also conducted a subjective picture quality study to further justify the perceptual relevance of our results.
\end{itemize}
The rest of this paper is organized as follows. Section II reviews related work. Section III unfold the details of our proposed resizing framework for learning-based compression models. Experiments and analysis are given in Section IV. Finally, Section V concludes the paper and draws future possible directions of this research.

\section{Background}
\subsection{Learning-based Lossy Image and Video Compression}
Recent years have witnessed a great surge of invention in the design of lossy image compression models realized by deep autoencoder architectures. They have been shown to achieve performances competitive with classical image coding standards, including JPEG2000 and HEVC Intra. Early efforts \cite{BalleLS16a,TheisCAE17} focused on tackling fundamental challenges like non-differential quantization in order to construct end-to-end trainable infrastructures. Unlike other image-to-image transformation models, which mainly focus on reducing distortion, the bitrate is also approximated and taken into account during training. Later, more sophisticated designs were explored to extract increasingly compact features, or to reconstruct images with higher quality from compressed latent representations. For example, some recent approaches have adopted recurrent neural networks (RNNs) \cite{Toderici2015VariableRI,Toderici2017,Johnston_2018_CVPR} to recursively compress residual information. Generative adversarial network (GANs) based compression models \cite{agustsson2019generative,Lhdefink2019GANVJ,mentzer2020high} produce reconstructed images whose statistics resemble the ground truth distributions. Although they often do not perform well with respect to pointwise quality measures, they can produce perceptually pleasing outcomes, especially at very low bitrates. Other methods like multi-scale networks \cite{Nakanishi2019} and invertible wavelet structures \cite{MaIWave2020} have also been explored.

Another popular research direction has focussed on improving entropy estimation, which directly affects the bits that are required in rate-distortion optimization. In \cite{balle2018variational}, a scale hyperprior is introduced into the compression model. The authors use an additional network to estimate the standard deviation, to better model the conditional probability of the latent representation. Minnen \textit{et al.} \cite{NIPS2018_8275} and Lee \textit{et al.} \cite{Lee2019Context} further incorporate context-adaptive models to reduce local redundancies. These context models are often realized by PixelCNN-like modules. Other solutions have used 3D-CNNs to facilitate conditional probability modeling \cite{Mentzer_2018_CVPR,ChenContext2021}, or have adopted Gaussian mixture models \cite{ChengGMM_2020_CVPR} to improve likelihood estimation.

Beyond still pictures, considerable progress has also been made extending these ideas to video compression. Early attempts on end-to-end video compression, such as \cite{wu2018vcii,cheng19}, have employed \textit{frame interpolation} schemes which temporally interpolate frames in a video using neural networks. Residuals between source and interpolated frames are then encoded. Motivated by the motion estimation and motion compensation (MEMC) scheme in hybrid video codecs, a series of \textit{optical flow}-based methods have also been proposed. For example, the DVC model \cite{LuDVC2019CVPR} uses a pretrained optical flow model to provide temporal predictions. Rippel \textit{et al.} \cite{rippel2019iccv} generalized the motion estimation process to achieve more efficient latent representations, and to compress motion and residuals simultaneously with a single autoencoder. In \cite{djelouah12019iccv}, a pretrained optical flow estimation model was combined with a learned encoder-decoder pair to perform bidirectional inter frame interpolation. Other variants of the optical flow setting, such as modeling motion using scale-space flow \cite{agustsson2020CVPR} or multi-scale flow \cite{liuNVC2021}, have been proposed to obtain promising improvements in coding efficiency. Interestingly, instead of estimating and signaling motion explicitly, the MOVI-Codec \cite{MXChenPCS2021} captures motion regularities from displaced frame differences, without conducting any motion search.

\begin{figure}[!t]
	\centering
	\footnotesize
	\renewcommand{\tabcolsep}{0pt} 
	\renewcommand{\arraystretch}{1.1} 
	\def\imgwid{}
	\begin{tabular}{c}
    \includegraphics[width=0.47\textwidth]{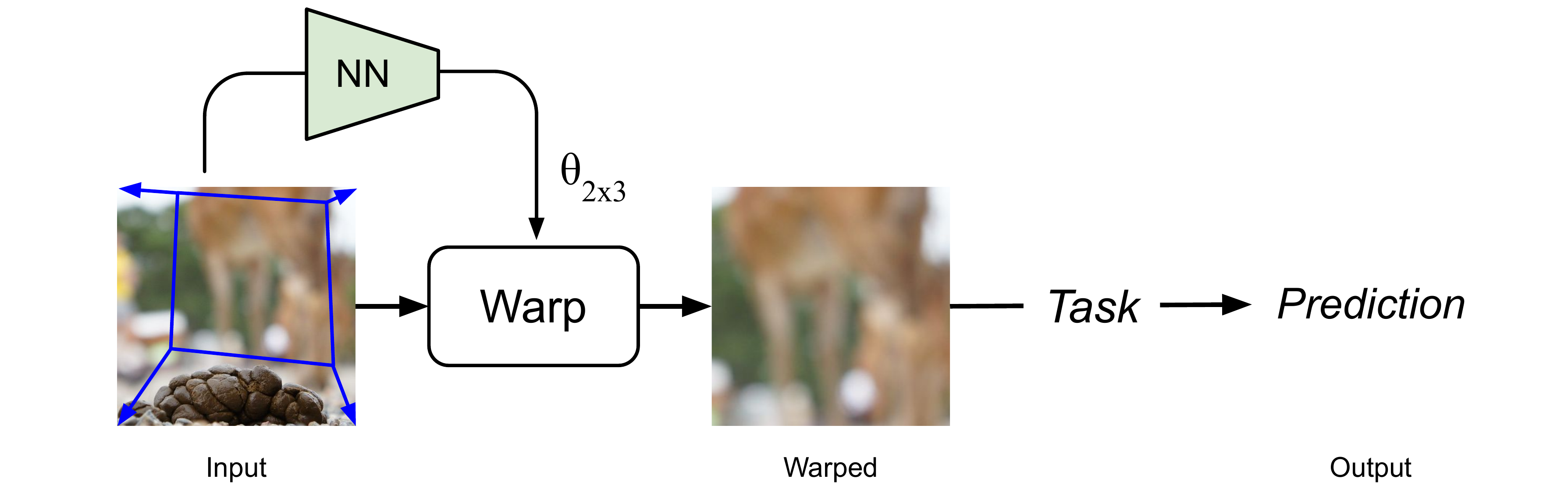} \\
    (a)\\
    \includegraphics[width=0.47\textwidth]{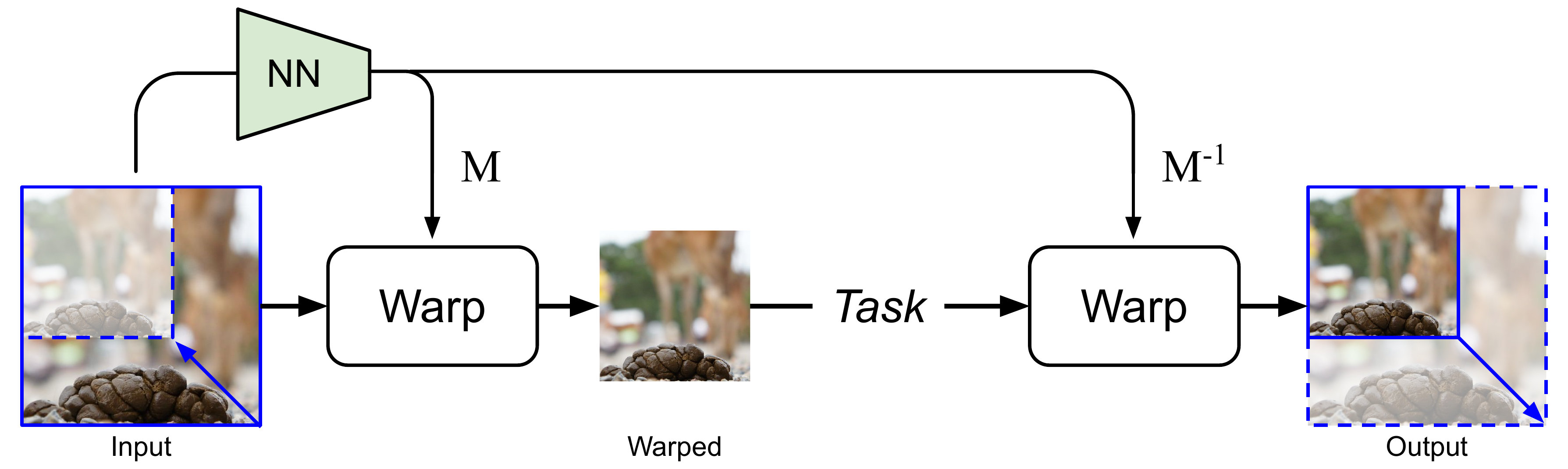} \\
    (b)\\
	\end{tabular}
	\caption{Comparison of (a) the spatial transformer network (STN) \cite{NIPS2015_33ceb07b} and (b) our framework of a pair of resizing layers for the compression task. The solid blue boxes indicate the warp of the input.}
	\label{fig:compare_stn}
\end{figure}

\begin{figure*}[!ht]
	\centering
	\footnotesize
	\renewcommand{\tabcolsep}{0pt} 
	\renewcommand{\arraystretch}{1.5} 
	\begin{tabular}{c}
    \includegraphics[width=0.98\textwidth]{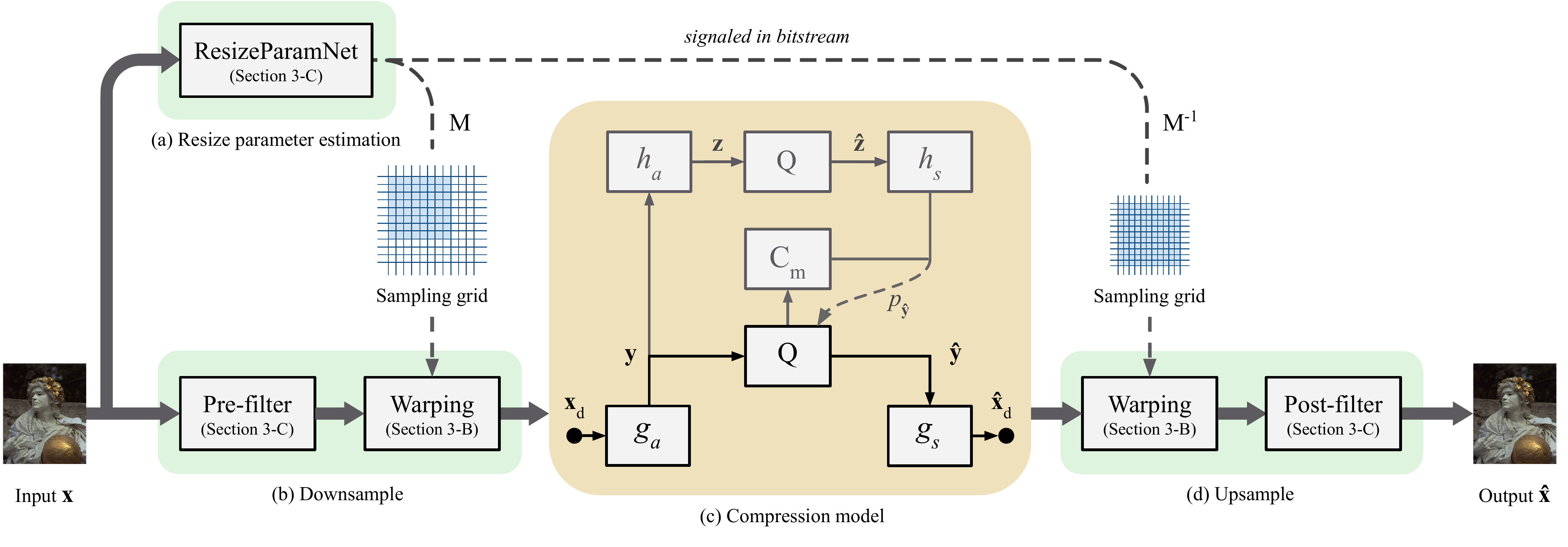}
	\end{tabular}
	\caption{\textbf{Overview of our proposed \rc.} First (a) estimate the resize factor $M$ given an input source image. Using $M$, the image is downscaled and upscaled by module (b) and (d), respectively. Blocks shadowed in green are new components. Bold arrows indicate the flow of data in the framework, while thin dashed arrows represent the control signals being delivered to the resizing modules.}
	\label{fig:overview}
\end{figure*}

Unlike the approaches mentioned above, our framework is a straightforward resizing-based compression framework that can be applied on any kind of neural compression kernels. In fact, the idea behind our resize parameter estimator is conceptually inspired by Spatial Transformer Networks (STN) \cite{NIPS2015_33ceb07b}, which is a type of neural network that is able to learn spatial affine transform parameters between images. Despite sharing some similarities, we point out important differences between our approach and STNs in Fig. \ref{fig:compare_stn}. The STN predicts a $2\times3$ affine transform matrix $\theta_{2\times3}$ that is used in a bilinear warping layer. Since STNs are commonly used in high-level computer vision tasks like image recognition, it often results in cropping out a trapezoidal region of interest from the input image. However, when applying this concept to the compression task, a scalar $M$ is instead learned: the resize parameter. To ensure that the reconstruction has the same resolution as the input, another warping layer is constructed, and constrained as an inverse transform back to the original geometry.

\subsection{Resizing and its Application in Image and Video Coding}
Unsurprisingly, the idea of applying spatial resizing in the context of conventional codecs has been deeply investigated and implemented in widespread practice. Typical use cases can be roughly categorized into the two classes. The first class includes codec-agnostic resizers operating in an \textit{out-of-loop} manner, whereby a high-resolution source frame is spatially downscaled before encoding. An upscaling algorithm is implemented on the decoder side to scale the reconstructed frame back to its original resolution. Studies on image codecs like JPEG and JPEG2000 \cite{Bruckstein2003,Lin2006,XiaolinWu2009}, and on video compression \cite{Knoche2005,Cermak2011,Georgis2016} have shown that encoding at lower resolutions generally results in better quality, when compressing to low bitrates. These schemes balance distortions produced by scaling against those by compression. Also in this class is a method often employed by streaming companies \cite{cvx_pertitle15, CChen2018, Wu2020}, where each source video is encoded at a finite set of combinations of resolution and compression levels, yielding multiple rate-distortion (RD) curves. Then, optimal encoding recipes are selected on the convex hull of the RD curves. This concept is simple, yet it operates at the extreme expense of exhaustive search, which is energy intensive and is also hard to apply in real-time applications. In efforts to avoid the need for search, more sophisticated resolution adaptation approaches have also been investigated \cite{Toni2015, Li2016, Sani2017, Afonso2019, Zhang19ViSTRA2, Bhat2020}.

The second class includes resizing incorporated as a normative \textit{in-loop} coding tool. Shen \textit{et al.} \cite{MinminShen2011} proposed to encode inter frames at reduced resolutions. An example-based super resolution algorithm was designed to reconstruct high quality frames. The AV1 codec standard describes an option to horizontally scale a source frame to a lower resolution. Before updating reference frame buffers, linear upsampling is employed as part of an in-loop restoration filter \cite{JoshiAV1SR2019,HanAV12021}. In the standardization of VVC, an adaptive resolution change (ARC) scheme is defined \cite{JVETM0135} that enables inter prediction between frames having different resolutions. Improvements have been made on corresponding coding tools \cite{ChangARC2020,FuRPR2021} to further improve coding efficiency.

\section{Proposed Method}\label{sec:proposed_method}
Next we provide an overview of the Resize-Compress framework. We introduce the design methodology for each component, and analyze the learned resize factor both quantitatively and qualitatively. Finally, we present details on training and implementation.

\subsection{Overview}\label{sec:overview}
Our proposed end-to-end trainable framework for resizing and learned compression is depicted in Fig. \ref{fig:overview}. Given an uncompressed source image $\mathbf{x}$, the \resizenet~network first maps $\mathbf{x}$ into a parameter $M$, the resize factor for the downsample module in Fig. \ref{fig:overview}(b). Accordingly, the estimated parameter $M$ is used to generate a sampling grid $\mathcal{T}_M$, which is a set of fractional coordinates where the input should be sampled to produce the resampled output. To mitigate the information loss introduced by subsampling, a learnable convolutional layer (Pre-filter) is placed beforehand, allowing the retention of information. This is quite analogous to the precoding technique \cite{Albreem2021} used in communication systems, where channel information is coded on the input signal.

The image that has been downscaled by the warping layer, denoted by $\mathbf{x}_\text{d}$, is then encoded by a deep compression model. A general compression system (Fig. \ref{fig:overview}(c)) comprises an analysis transform $g_a$ at the encoder side, and a synthesis transform $g_s$ at the decoder side. The latent representation $\mathbf{y}$ is first generated by applying the transformation $g_a$ on the input
\begin{equation}
  \mathbf{y}=g_a\left(\mathbf{x}_\text{d}\right).
\end{equation}
Then, $\mathbf{y}$ is quantized by the module $\text{Q}$ and synthesized back by $g_s$, yielding the reconstructed image
\begin{equation}
  \hat{\mathbf{x}}_\text{d}=g_s\left(\hat{\mathbf{y}}\right)=g_s\left(\text{Q}\left(\mathbf{y}\right)\right).
\end{equation}
Other modules such as a hyperprior ($h_a$, $h_s$) \cite{balle2018variational} or context model ($C_m$) \cite{NIPS2018_8275,Lee2019Context} could also be utilized to obtain better estimates of the parameterized probability distribution. It is worth mentioning that the quantized representations (e.g., $\hat{\mathbf{y}}$ and $\hat{\mathbf{z}}$) are encoded as discrete-valued data into the bitstream using an arithmetic coder. Finally, $\hat{\mathbf{x}}$ is output by upsampling $\hat{\mathbf{x}}_\text{d}$ using a similar warping layer (Fig. \ref{fig:overview}(d)), but with a reciprocal scaling parameter $M^{-1}$. Unlike the downsampling module in Fig. \ref{fig:overview}(b), the post-filtering network is placed \textit{after} the warping layer to extract features that can repair the spatially degraded image. Similar methods have worked well in image super-resolution architectures \cite{Dong2016}.

\subsection{Differentiable Resize Layer}\label{sec:diff_resize_layer}
It has been shown \cite{NIPS2015_33ceb07b} that operations involving backward image warping with interpolation kernel $k(\cdot)$ are (sub-)differentiable with respect to all the arguments, allowing for the gradients to be back-propagated through the forward model. Inspired by this, we implement our resizing layers as a constrained form of image warping. We follow the convention in \cite{NIPS2015_33ceb07b} to define a parameterized sampling grid $\mathcal{T}_M$ for the $i
$th pixel, which is a spatial affine coordinate transformation on the target coordinates of the warping output $(x^t_i,y^t_i)^\intercal$
\begin{equation}\label{eq:grid}
  \mathcal{T}_M=
  \begin{pmatrix}
    x^s_i\\
    y^s_i
  \end{pmatrix}
  =
  \begin{bmatrix}
    \frac{1}{M} & 0 \\
    0 & \frac{1}{M} 
  \end{bmatrix}
  \begin{pmatrix}
    x^t_i\\
    y^t_i
  \end{pmatrix}
  \\
  ~~\forall i \in [1,...,HW],
\end{equation}
where $(x^s_i,y^s_i)^\intercal$ denote the corresponding input coordinates that define the sample points for interpolation. Under this scaling-constrained transformation, resolution is reduced when $0\le M<1$, while $M>1$ means an upscaling operation is performed. The output value of a particular pixel $V_i$ located at $(x^t_i,y^t_i)^\intercal$ can be written as
\begin{equation}\label{eq:warp}
  V_i=\sum_n\sum_m U_{mn} k(x^s_i-m)k(y^s_i-n),
\end{equation}
where $U_{mn}$ is the input pixel value at location $(n,m)^\intercal$. Therefore, warping an image $I=[U_{i}]$ can be expressed by
\begin{equation}\label{eq:warp2}
  \text{Warp}_{\mathcal{T}_M,k}\left(I\right)=\text{Warp}_{\mathcal{T}_M,k}\left([U_i]\right)=\left[V_i\right].
\end{equation}
Note that an unwanted boundary could be created during downsampling, when the resampling grid $(x^s_i,y^s_i)^\intercal$ exceeds the dimensions of the input source. In our implementation, we simply remove the unwanted boundary by cropping the warped image to $S\left\lceil\frac{\text{width}}{S}\right\rceil \times S\left\lceil\frac{\text{height}}{S}\right\rceil$, which is divisible by the equivalent stride of the compression model $S$.

In order to properly select the interpolation kernel $k$ for our problem, we empirically conducted the following simplified tasks:
\begin{enumerate}
  \item \textbf{Scale down:} An input $x=I$ is warped by a factor of $M$ to fit the target image $y=\text{Warp}_{\mathcal{T}_{N},k}\left(I\right)$, $N=0.5$
  \begin{equation}\label{eq:task1}
    f_1(x)=\text{Warp}_{\mathcal{T}_M,k}\left(x\right).
  \end{equation}
  \item \textbf{Scale up:} A downscaled input $x=\text{Warp}_{\mathcal{T}_{N},k}\left(I\right)$, $N=0.5$ is warped by a factor of $M^{-1}$ to fit the target image $y=I$
  \begin{equation}\label{eq:task2}
    f_2(x)=\text{Warp}_{\mathcal{T}_{M^{-1}},k}\left(x\right).
  \end{equation}
  \item \textbf{A resize pair:} An input $x$ warped (downscaled) by a factor of $M$ followed by an inverse warping ($x=y=I$)
  \begin{equation}\label{eq:task3}
    f_3(x)=\text{Warp}_{\mathcal{T}_{M^{-1}},k}\left(\text{Warp}_{\mathcal{T}_{M},k}\left(x\right)\right).
  \end{equation}
\end{enumerate}

\begin{figure}[!t]
	\centering
	\footnotesize
	\renewcommand{\tabcolsep}{0pt} 
	\renewcommand{\arraystretch}{1.6} 
	\def\imgwid{}
	\begin{tabular}{l}
    ~~~~~Task 1: $I \xrightarrow{\downarrow M} \text{Warp}_{\mathcal{T}_M,k}\left(I\right) \xRightarrow{\text{~fit~~}} \text{Warp}_{\mathcal{T}_{.5},k}\left(I\right)$\\
    \includegraphics[width=0.46\textwidth]{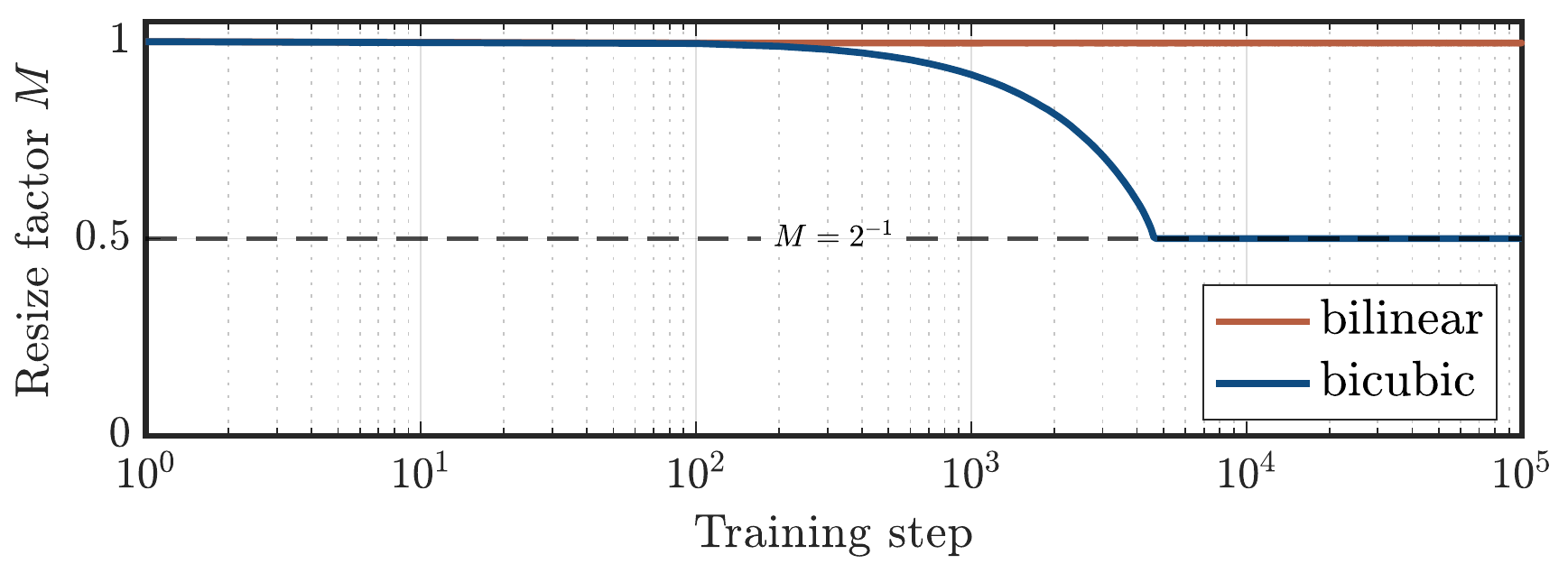} \\
    ~~~~~Task 2: $\text{Warp}_{\mathcal{T}_{.5},k}\left(I\right) \xrightarrow{\uparrow M} \text{Warp}_{\mathcal{T}_{M^{-1}},k}\left(\text{Warp}_{\mathcal{T}_{.5},k}\left(I\right)\right) \xRightarrow{\text{~fit~~}} I$\\
    \includegraphics[width=0.46\textwidth]{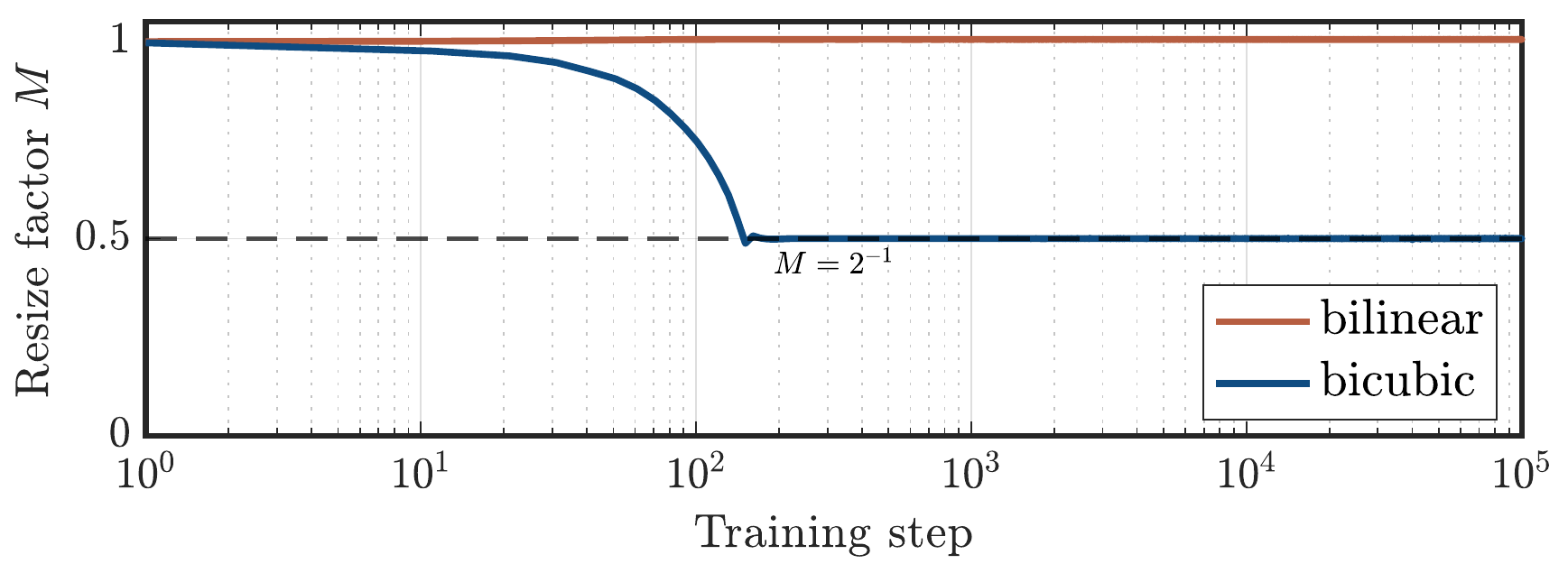} \\
    ~~~~~Task 3: $I \xrightarrow{\downarrow M~\uparrow M} \text{Warp}_{\mathcal{T}_{M^{-1}},k}\left(\text{Warp}_{\mathcal{T}_M,k}\left(I\right)\right) \xRightarrow{\text{~fit~~}} I$\\
    \includegraphics[width=0.46\textwidth]{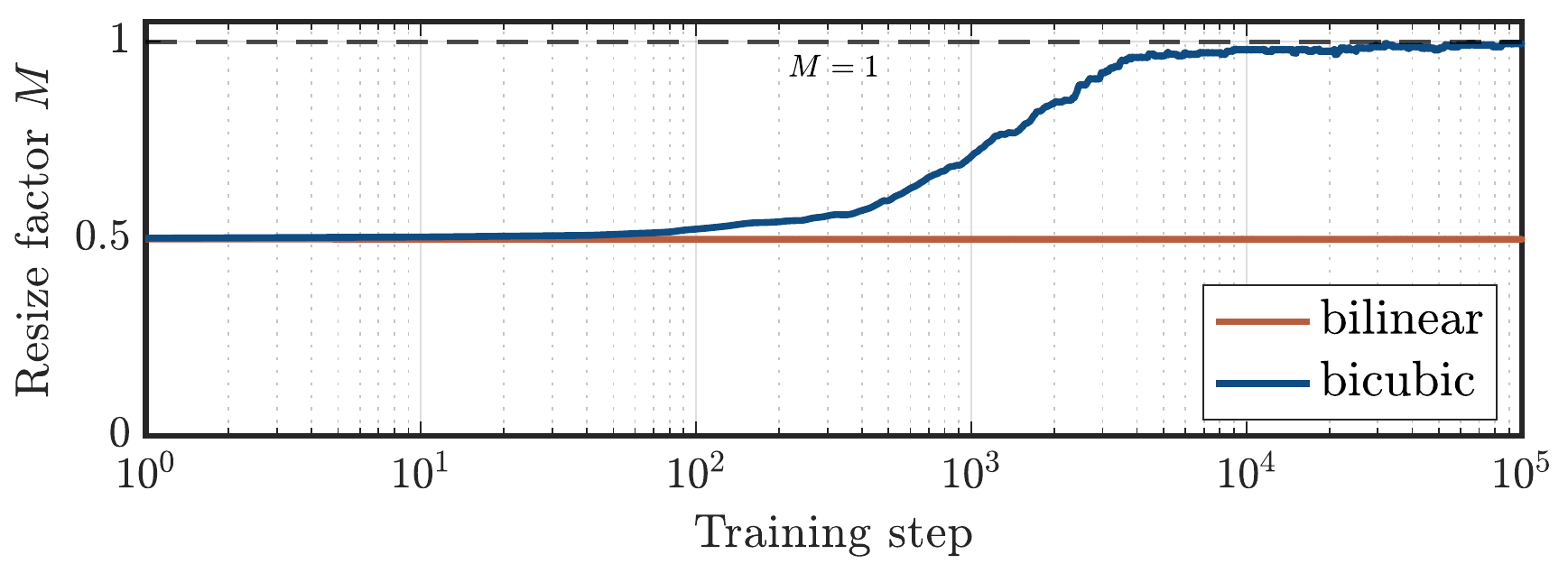} \\
	\end{tabular}
	\caption{Training plot of the predicted resize factor $M$ using bilinear and bicubic interpolation kernels in different tasks. The black dashed lines denote the optimal values of $M$ in each task as in (\ref{eq:task_opt}).}
	\label{fig:compare_interpolation}
\end{figure}

Note that $N$ is a constant while $M$ is a trainable variable, rather than the output of \resizenet. For (\ref{eq:task1})--(\ref{eq:task3}), we optimize $M$ such that the MSE between $f_{i}(x)$ and $y$ is minimized. Ideally, the value of $M$ should converge to
\begin{equation}\label{eq:task_opt}
\operatorname*{arg\,min}_M \norm{f(x)-y}_2=
\left\{
	\begin{array}{ll}
		2^{-1},  & \mbox{if } f(x)=f_1(x) \\
    2^{-1},& \mbox{if } f(x)=f_2(x) \\
		1,  & \mbox{if } f(x)=f_3(x)
	\end{array}
\right..
\end{equation}
These optimal parameters are intuitively determined for each task. For example, $M=1$ minimizes the error for the third task defined in (\ref{eq:task3}), since it is equivalent to an identity transformation, preserving the fidelity of the input. Figure \ref{fig:compare_interpolation} compares the parameter $M$ recorded during training with respect to the use of bilinear and bicubic interpolation. It may be observed that $M$ quickly approached the optimal values in (\ref{eq:task_opt}) when bicubic interpolation was used, whereas bilinear interpolation did not yield good convergence. This is likely due to the larger receptive field of the bicubic kernel, which results in stronger gradient signals for back propagation. Motivated by these observations, we implemented \textit{bicubic} warping in our framework.

\begin{table}[!t]
  \renewcommand{\arraystretch}{1.6}
  \renewcommand{\tabcolsep}{3.2pt} 
  \caption{Architectural details of the additional networks in our framework.}
  \label{tab:network_architecture}
  \centering
  \begin{tabular}{c | l l l l l}
  & Name & Input & Operation & Activation & Output dim. \\
  \hline
  \multirow{5}{*}{\rotatebox{90}{\footnotesize\textbf{Pre/Post-filter}}}
  & input  & ---   & ---      & ---                       & $W\times H\times 3$\\
  & C1     & input & Conv: 3$\times$3$\,\vert\,$c32$\,\vert\,$s1 & ReLU      & $W\times H\times 32$\\
  & C2     & C1    & Conv: 3$\times$3$\,\vert\,$c32$\,\vert\,$s1 & ReLU      & $W\times H\times 32$\\
  & C3     & C2    & Conv: 3$\times$3$\,\vert\,$c3$\,\vert\,$s1 & tanh      & $W\times H\times 3$\\
  & output & input, C3 & Addition & ---                  & $W\times H\times 3$\\
  \hline
  \hline
  \multirow{6}{*}{\rotatebox{90}{\footnotesize\textbf{\resizenet}}}
  & input  & ---   & ---      & ---                       & $W\times H\times 3$\\
  & R1     & input & Res: $C=16$  & ReLU  & $\frac{W}{2}\times\frac{H}{2}\times 16$\\
  & R2     & R1    & Res: $C=32$  & ReLU  & $\frac{W}{4}\times\frac{H}{4}\times 32$\\
  & R3     & R2    & Res: $C=64$  & ReLU  & $\frac{W}{8}\times\frac{H}{8}\times 64$\\
  & P4     & R3    & Global Avg. Pooling      & ---   & $1\times 1\times 64$  \\
  & output & P4    & Mean                     & ReLU  & 1   \\
  \hline
  \multirow{8}{*}{\rotatebox{90}{\footnotesize\textbf{Residual Block} (Res)}}
  & input  & ---   & parameters: $C$          & ---  & $W\times H\times C_{in}$\\
  & C1     & input & Conv: 3$\times$3$\,\vert\,$c$C$$\,\vert\,$s1 & ---  & $W\times H\times C$\\
  & BN1    & C1    & Batch Norm.              & ReLU & $W\times H\times C$\\
  & C2     & BN1   & Conv: 3$\times$3$\,\vert\,$c$C$$\,\vert\,$s1 & ---  & $W\times H\times C$\\
  & BN2    & C2    & Batch Norm.              & ReLU & $W\times H\times C$\\
  & C3     & input & Conv: 3$\times$3$\,\vert\,$c$C$$\,\vert\,$s1 & ---  & $W\times H\times C$\\
  & A4     & BN2, C3 & Addition               & ---  & $W\times H\times C$\\
  & output & A4    & MaxPool: 2$\times$2$\,\vert\,$s2          & ReLU & $\frac{W}{2}\times\frac{H}{2}\times C$  \\
  \hline
  \end{tabular}
  \vspace{1ex}

  {\raggedright Conv: convolutional layers denoted as kernel size$\,\vert\,$\# of channels$\,\vert\,$stride.\par}
  {\raggedright MaxPool: max pooling layers denoted as pooling size$\,\vert\,$
  stride.\par}
  {\raggedright $C_{in}$: Number of input channels.\par}
\end{table}

\subsection{Network Architecture}
The details of the networks are outlined in Table \ref{tab:network_architecture}, including the auxiliary module that is used to estimate resize factors, as well as the shallow filter networks placed at the two ends of the compression system.

\textbf{The \resizenet:} The goal is to learn an $\mathbb{R}^{H\times W\times 3}\mapsto \mathbb{R}^{1}$ transformation that maps an input $\mathbf{x}$ to a resize factor $M$ via a CNN. We constructed a network with similar design as the standard ResNet \cite{HeZRS16}, consisting of three stages of residual blocks. The spatial size is reduced by a factor of $2$ after each stage via $2\times 2$ max pooling layers. Finally, $64$ feature maps are fed to a global average pooling (GAP) layer, and the output is obtained by averaging the $64$ values. The parameterization of each layer is detailed in the table. We experimented with different ways, such as using a fully connected (FC) layer, to aggregate the feature maps. However, we did not obtain improvements in performance.

\textbf{The pre/post-filter networks:} Both the \textit{pre-filter} and the \textit{post-filter} share the same network architecture of three stages of convolutional layers, accepting a $3$-channel signal as input. The sizes of the convolutional layers are all fixed at $3\times3$, while the number of filters is $32$. We zero pad the boundaries of the feature maps before applying each convolution, so that the output size is not reduced. Except for the last layer, all of the convolutional layers are activated by a ReLU nonlinearity. Finally, a $3$-channel output is produced, yielding a residual that is added element-wise to the input image. It should be noted that the resolution changes produced by our model are handled by individual resize layers as described in Section \ref{sec:diff_resize_layer}, hence the stride parameter is always set to $s=1$. We note that the two filters do not share parameters.

\begin{figure*}[!t]
	\centering
	\footnotesize
	\renewcommand{\tabcolsep}{0.9pt} 
	\renewcommand{\arraystretch}{0.7} 
	\def\imgwid{0.09\textwidth}
	\begin{tabular}{c ccc c ccc c ccc}
    & \multicolumn{11}{c}{Reduced~$\xleftarrow{\hspace*{5.7cm}}$~~\textbf{Resolution}~~$\xrightarrow{\hspace*{5.7cm}}$~~~~\,~~Full} \\
    \multirow{1}{*}{\rotatebox{90}{\text{Large~}$\xleftarrow{\hspace*{1.6cm}}$~~\textbf{Resolution}~~$\xrightarrow{\hspace*{1.6cm}}$\text{~Small}~~}}~~  
    \\
    &
    \includegraphics[width=\imgwid]{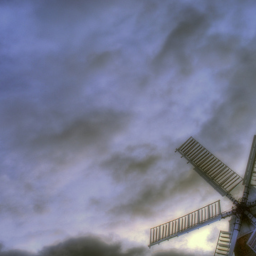} &
    \includegraphics[width=\imgwid]{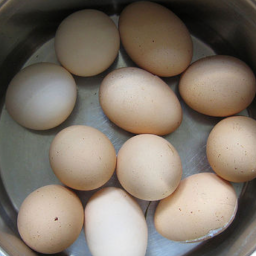} &
    \includegraphics[width=\imgwid]{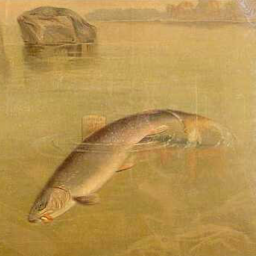} &~~~&
    \includegraphics[width=\imgwid]{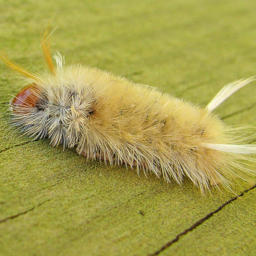} &
    \includegraphics[width=\imgwid]{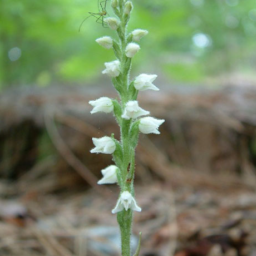} &
    \includegraphics[width=\imgwid]{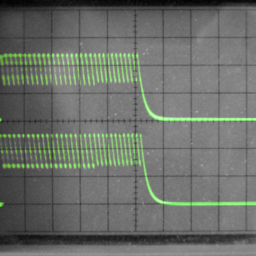} &~~~&
    \includegraphics[width=\imgwid]{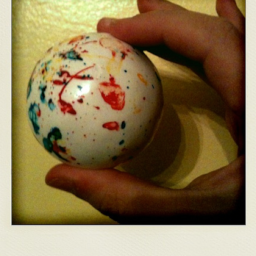} &
    \includegraphics[width=\imgwid]{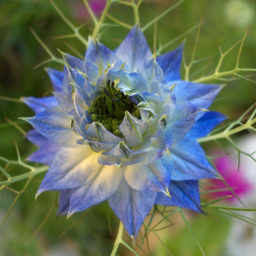} &
    \includegraphics[width=\imgwid]{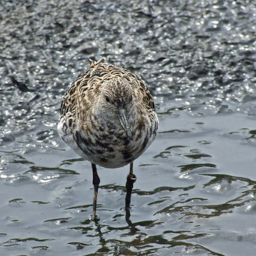} \\
    &
    \includegraphics[width=\imgwid]{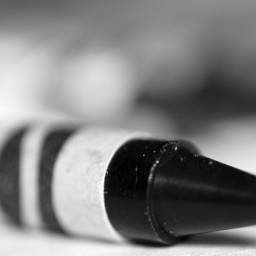} &
    \includegraphics[width=\imgwid]{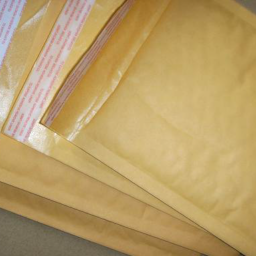} &
    \includegraphics[width=\imgwid]{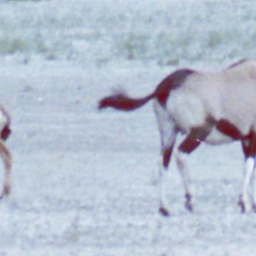} &~&
    \includegraphics[width=\imgwid]{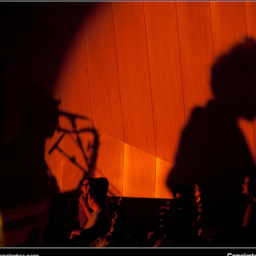} &
    \includegraphics[width=\imgwid]{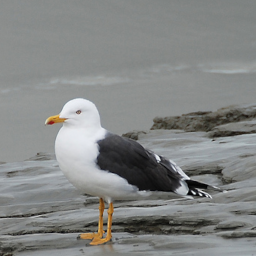} &
    \includegraphics[width=\imgwid]{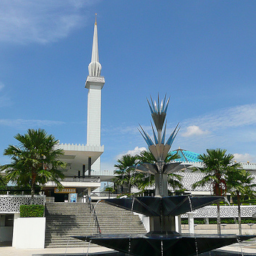} &~&
    \includegraphics[width=\imgwid]{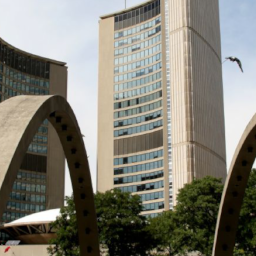} &
    \includegraphics[width=\imgwid]{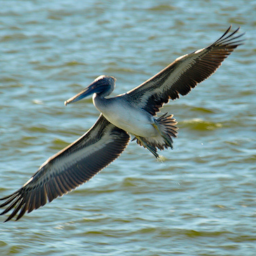} &
    \includegraphics[width=\imgwid]{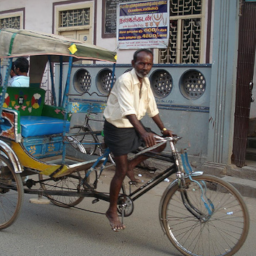} \\
    &
    \includegraphics[width=\imgwid]{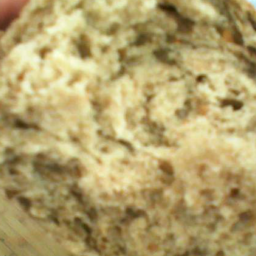} &
    \includegraphics[width=\imgwid]{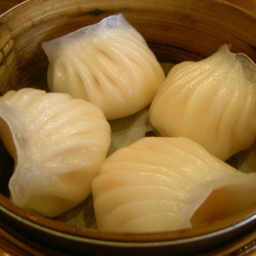} &
    \includegraphics[width=\imgwid]{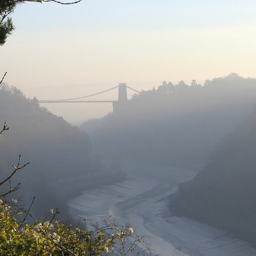} &~&
    \includegraphics[width=\imgwid]{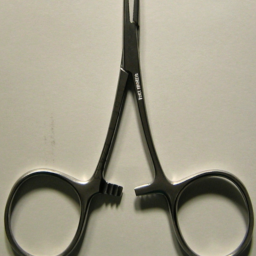} &
    \includegraphics[width=\imgwid]{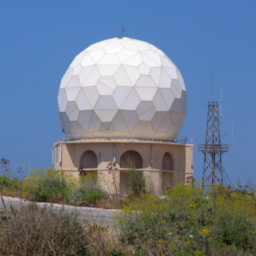} &
    \includegraphics[width=\imgwid]{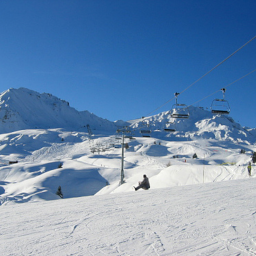} &~&
    \includegraphics[width=\imgwid]{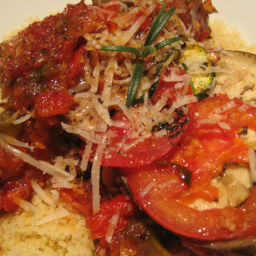} &
    \includegraphics[width=\imgwid]{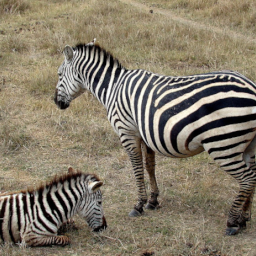} &
    \includegraphics[width=\imgwid]{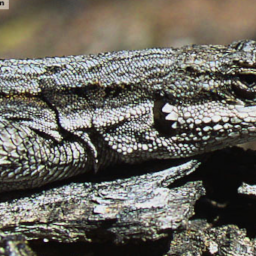} \\
    &
    \includegraphics[width=\imgwid]{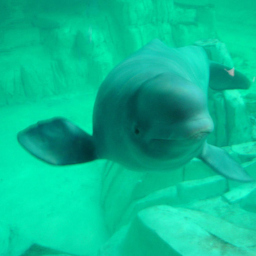} &
    \includegraphics[width=\imgwid]{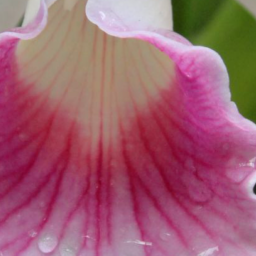} &
    \includegraphics[width=\imgwid]{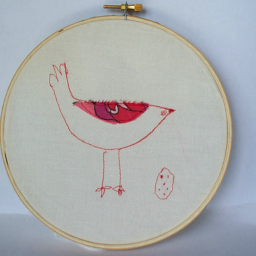} &~&
    \includegraphics[width=\imgwid]{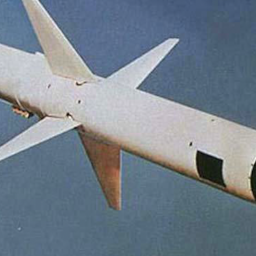} &
    \includegraphics[width=\imgwid]{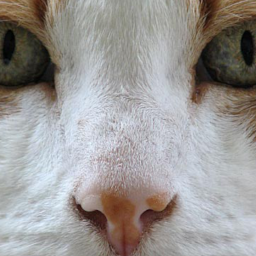} &
    \includegraphics[width=\imgwid]{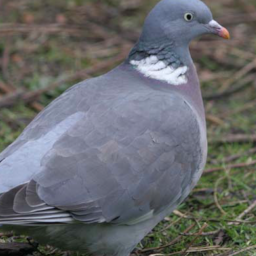} &~&
    \includegraphics[width=\imgwid]{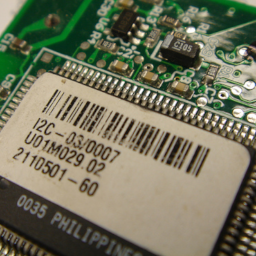} &
    \includegraphics[width=\imgwid]{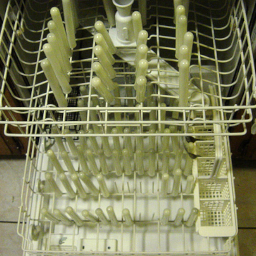} &
    \includegraphics[width=\imgwid]{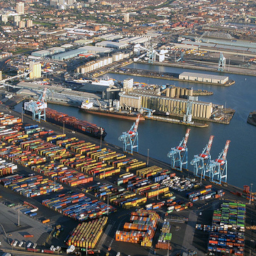} \\\\
    &
    \multicolumn{3}{c}{(a) $0<{M}\le0.33$} & ~\, & \multicolumn{3}{c}{(b) $0.33<{M}\le0.67$} & ~\, & \multicolumn{3}{c}{(c) $0.67<{M}\le1$} \\
	\end{tabular}
	\caption{Visualization of uncompressed \textit{validation} patches associated with different resize factors $M$ predicted by \resizenet. \resizenet~was trained with the Ball\'{e}17 model \cite{BalleLS16a} with $\lambda=0.001$. These example patches were extracted from ImageNet.}
	\label{fig:visual_factor}
\end{figure*}

\begin{figure}[!t]
  \centering
  \footnotesize
  \includegraphics[width=3.45in]{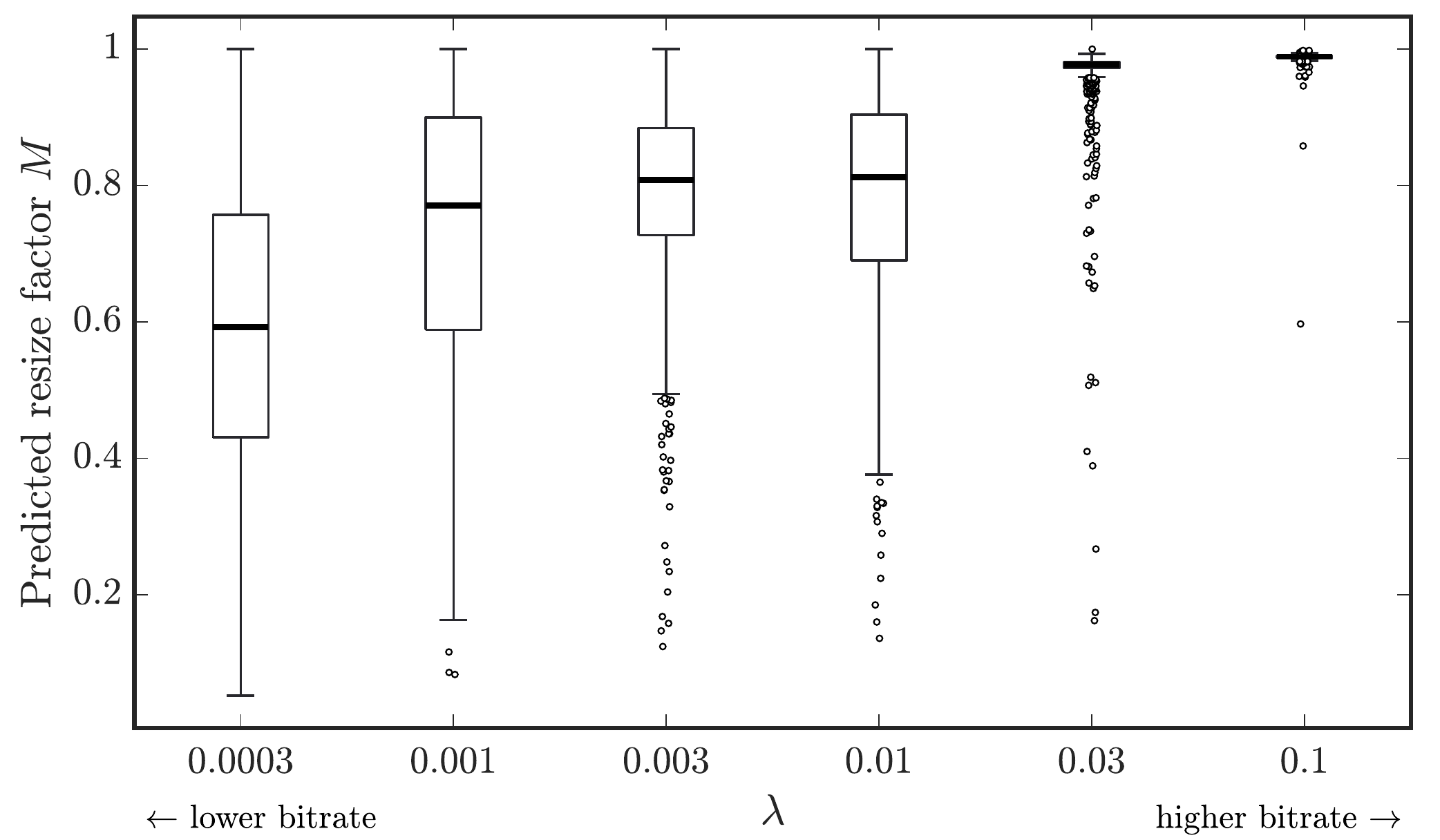}
  \caption{Box plot of predicted resize factor $M$ against the weighting parameter $\lambda$, evaluated on the \textit{validation} patches extracted from ImageNet. The `$\circ$' symbol denotes outliers. \resizenet~was trained with the Ball\'{e}17 model \cite{BalleLS16a}.}
  \label{fig:hist_res}
\end{figure}

\subsection{Loss Function}
Let $\theta_{g_a}$, $\theta_{g_s}$, $\phi_{\text{pre}}$, $\phi_{\text{post}}$, and $\phi_{\text{resize}}$ be the parameters of the analysis transform $g_a$, the synthesis transform $g_s$, the pre-filter, the post-filter, and \resizenet, respectively. Our goal is to end-to-end optimize these parameters, such that the pipeline can generate a reconstructed batch $\hat{\mathbf{x}}$ that has high fidelity relative to the source batch $\mathbf{x}$. Meanwhile, the cost of encoding quantized representations to bits should be as small as possible. Therefore, we train the model against the following losses. First, the distortion term measures the fidelity between the reconstructed image and its pristine source, which is defined as the residual between $\mathbf{x}$ and $\hat{\mathbf{x}}$ mapped by a distortion function $d$:
\begin{equation}\label{pixel_loss}
  \mathcal{L}_\text{dist}=d\left(\mathbf{x}-\hat{\mathbf{x}}\right).
\end{equation}
Here, the squared Euclidean distance $d(x)=\norm{x}_2^2$ is adopted to minimize the mean squared error (MSE).

To representing the bit consumption of quantized latents, the rate loss is defined by
\begin{equation}\label{rate_loss}
  \mathcal{L}_\text{rate}=\sum_{\hat{q}\in\mathcal{Q}}-\log_2 p_{\hat{q}}\left(\hat{q}\right),
\end{equation}
where $p_{\hat{q}}$ refers to an entropy model estimated over the unknown distribution of input images, and $\mathcal{Q}$ denotes the set containing all the respective quantized latents. For example, in the work of Ball\'{e} \textit{et al.} \cite{balle2018variational}, the quantized latents $\hat{\mathbf{y}}$ and hyper-latents $\hat{\mathbf{z}}$ are encoded into bits, which is a case where $\mathcal{Q}=\{\hat{\mathbf{y}}, \hat{\mathbf{z}}\}$. Coupling the losses from (\ref{pixel_loss}) and (\ref{rate_loss}), and all the trainable parameters collectively denoted by $\boldsymbol\theta=\{\theta_{g_a}, \theta_{g_s}, \phi_{\text{pre}}, \phi_{\text{post}}, \phi_{\text{resize}}, p_{\hat{q}\vert\hat{q}\in\mathcal{Q}}\}$, the overall training objective that is used to optimize the system is defined as:
\begin{equation}\label{eq:total}
  \mathcal{L}_\text{total}\left(\boldsymbol\theta\right) =\lambda\mathcal{L}_\text{dist}+\mathcal{L}_\text{rate},
\end{equation}
where $\lambda$ is a weight parameter that balances bitrate against distortion of the encoded bitstream. By increasing $\lambda$, better quality of reconstructed images can be achieved at the cost of compression ratio. The loss derivative was used to update the model parameters $\boldsymbol\theta$ by back-propagating through the feed forward model. It is also interesting to note that (\ref{eq:total}) does not contain any loss terms specific to the resize factor $M$. Instead, the networks learn to directly estimate the resize factor as a byproduct of minimizing the rate-distortion loss.

\subsection{Visualization of the Learned Resizing Factors} \label{sec:visualization}
To understand how a source image is resized in our framework, we used a subset of 1000 image patches from the ImageNet database, and fed them into \resizenet~trained with different weighting parameters $\lambda$. In Fig. \ref{fig:visual_factor}, exemplar patches are arranged according to the estimated resize factor $M$ to demonstrate the efficacy of \resizenet. Generally, contents containing significant detail and texture are assigned values of $M$ closer to $1$, indicating less resolution reduction will be applied before compression. This is not unexpected, since high-frequency components are more susceptible to degradation from scaling. On the other hand, the patches of less complexity, or that are more blurry, are compressed at lower resolutions using smaller values of $M$. This is because in such instances, bitrate consumption can be significantly reduced by downsampling, while still maintaining similar quality level.

Figure \ref{fig:hist_res} plots the spreads of the predicted resize factor for models trained with different values of $\lambda$. It is evident that the estimated resize factor $M$ increases with $\lambda$, as should be expected, since the loss function in (\ref{eq:total}) is dominated by the MSE term, making downsampling an undesirable operation. At the high-bitrate extreme of $\lambda=0.1$, the distribution of $M$ is centered around $1$. It is also worth noting that $M$ spans a wide range of values when $\lambda\le 0.001$. Similar to the observation we made in Fig. \ref{fig:visual_factor}, this indicates the complex interplay between picture content, bitrate, and distortions: under constrained bitrate conditions, there is room to improve the (perceptual) rate-distortion tradeoff, by optimally applying different amounts of image resizing.

\begin{table*}[!ht]
  \caption{Overall comparison of different codecs and results of optimized deep image compression on three datasets. Each cell shows the average change of BD-rate expressed as percent. The baseline comparison is against the Ball\'{e}17 model \cite{BalleLS16a}. Smaller or negative values indicate better coding efficiency.}

  \label{tab:overall}
  \centering
  \renewcommand{\arraystretch}{1.5}
  \renewcommand{\tabcolsep}{3.24pt} 
  \begin{tabular}{l rrrr c rrrr c rrrr}
  \toprule
    \multirow{1}{*}{Image dataset\pZ} & \multicolumn{4}{c}{\textsc{Kodak} \cite{kodak_data}} && \multicolumn{4}{c}{\textsc{Tecnick} \cite{stag_tecnick}} && \multicolumn{4}{c}{\textsc{JPEG AI} \cite{jpegai_data}}\\
    \cline{2-5} \cline{7-10} \cline{12-15}
    \multirow{1}{*}{BD-rate metric\pZ}
    & PSNR & \multicolumn{2}{c}{MS-SSIM~~~~~VIF} & VMAF && PSNR & \multicolumn{2}{c}{MS-SSIM~~~~~VIF} & VMAF &
    & PSNR & \multicolumn{2}{c}{MS-SSIM~~~~~VIF} & VMAF \\
  \hline
  JPEG & 
  +181.69 & +193.13 & +207.60 & \phantom{1}+90.27 &&
  +173.98 & +202.28 & +179.30 & +103.14 &&
  +167.29 & +184.31 & +192.59 & \phantom{1}+85.44 \\

  JPEG2000 & 
  -12.70 &  +0.07 &  +3.81 & -33.54 &&
  -13.95 &  -9.92 &  -9.02 & -34.20 &&
  -17.79 & -11.69 & -10.53 & -35.80 \\

  WebP & 
  -0.80  & +2.72 & +20.56 & -18.90 &&
  +11.24 & -2.67 & +18.21 & -15.78 &&
  +1.41  & -0.94 & +18.31 & -16.40 \\

  HEVC Intra & 
  -31.34  &  -7.44 &  -17.22 & -30.39 &&
  -31.07  & -13.79 &  -23.73 & -29.43 &&
  -31.52  & -12.01 &  -24.40 & -25.62 \\
  \rowcolor{lightgray!20}
  Ball\'{e}17 \cite{BalleLS16a} \scriptsize{(Baseline)}& 
  +0.00 & +0.00 & +0.00 & +0.00 &&
  +0.00 & +0.00 & +0.00 & +0.00 &&
  +0.00 & +0.00 & +0.00 & +0.00\\
  \rowcolor{lightgray!20}
  \textbf{Ball\'{e}17~+Resize} \scriptsize{(Ours)} & 
  -5.20  & -6.60  & -8.53  & -10.25 &&
  -9.74  & -14.40 & -14.62 & -12.91 &&
  -11.21 & -11.44 & -14.13 & -10.60 \\

  Ball\'{e}18-Fact \cite{balle2018variational} & 
  -11.52 & -3.92 & -2.29 & -12.53 &&
  -8.52 & -6.70 & -4.34 & -14.30 &&
  -9.94 & -4.62 & -4.31 & -12.24 \\

  \textbf{Ball\'{e}18-Fact~+Resize} \scriptsize{(Ours)} & 
  -11.46 & -7.56 & -9.60 & -17.84 &&
  -12.51 & -17.67 & -15.60 & -20.09 &&
  -14.15 & -15.12 & -16.00 & -19.03 \\
  \rowcolor{lightgray!20}
  Ball\'{e}18-Hyper \cite{balle2018variational} & 
  -25.81 & -13.07 & -12.87 & -31.08 &&
  -28.43 & -21.45 & -21.45 & -35.89 &&
  -28.11 & -19.51 & -17.25 & -33.52 \\
  \rowcolor{lightgray!20}
  \textbf{Ball\'{e}18-Hyper~+Resize} \scriptsize{(Ours)} & 
  -24.91 & -19.69 & -23.88 & -42.93 &&
  -31.69 & -32.44 & -33.74 & -45.81 &&
  -32.45 & -29.76 & -33.89 & -44.21 \\

  Cheng20 \cite{ChengGMM_2020_CVPR} & 
  -50.78 & -42.95 & -45.39 & -56.20 &&
  -50.87 & -48.75 & -47.19 & -58.28 &&
  -47.35 & -43.57 & -46.66 & -54.58 \\

  \textbf{Cheng20~+Resize} \scriptsize{(Ours)} & 
  -47.69 & -44.66 & -47.38 & -58.00 &&
  -50.34 & -49.45 & -51.20 & -59.30 &&
  -49.07 & -44.69 & -48.80 & -55.65 \\

  \bottomrule
\end{tabular}
\end{table*}

\subsection{Implementation and Training Details}
We developed and trained our models in Python using the TensorFlow framework (version 1.15). All of the models were trained using NVIDIA Tesla K80 GPU cards. We used the CLIC Professional Dataset \cite{clic_data}, an image dataset consisting of high quality pictures, as training data. We did not use images smaller than 384 pixels along either the vertical or horizontal axix, resulting in a subset of $1600$ images. No augmentation was applied on the training data.  

During training, we used the Adam solver \cite{kingma:adam} to optimize the networks, with parameters $\beta_1=0.9$, $\beta_2=0.999$ and a batch size of $8$ with a crop size of 256 pixels. The networks were initially trained over $1$M iterations, using a learning rate that was fixed at $1e^{-4}$. Then, the learning rates were decreased to $3e^{-5}$ for an additional $1$M iterations. We found that using larger patches tended to facilitate the training stability of our framework, but with much slower training speed. Thus, we further refined the models using patches of size 384$\times$384 for only $100$K steps, resulting in a total of $2.1$M iterations of backpropagation. For fair comparison, all of the models (including the original baselines) were trained from scratch under the same conditions.

\section{Experiments and Analysis}
\subsection{Evaluation Experiments}\label{evalexp}
\textbf{Evaluation Dataset} To compare our method and various image codecs, we utilized the well-known Kodak dataset \cite{kodak_data} of $24$ very high quality uncompressed 768$\times$512 images. This publicly available image set is commonly used to evaluate image processing algorithms. We also used the Tecnick dataset \cite{stag_tecnick} containing $100$ images of 1200$\times$1200 resolution, and $16$ test images that are nearly ultra high definition from the JPEG-AI Call for Evidence \cite{jpegai_data}, yielding images having more diverse resolutions and contents. None of the test images were included in the training sets, to avoid bias and overfitting problems.

\textbf{Evaluation Protocol} The coding efficiency on the test set was measured by the Bj\o{}ntegaard-Delta bitrate (BD-rate) \cite{BDRate01} of each image codec, which quantifies average differences in bitrate at a same distortion level relative to another reference encode. A negative BD-rate means that the bitrate was reduced as compared with the baseline. We encoded the images at $8$ approximate different bitrates, ranging from $0.05$ bpp (bits per pixel) to $1$ bpp. Then, the BD-rates with respect to a variety of quality models were calculated. Aside from measuring the pixel-wise PSNR for completeness, we mainly relied on perception-based quality models, including MS-SSIM \cite{WangMSSSIM03}, VIF \cite{SheikhB06} and VMAF \cite{ZliVMAF18}, to quantify the distortion levels that were used for BD-rate calculation. We are aware of recent studies \cite{JPEGAI19,Testolina2021} of quality evaluation on deep learning based image compression. It has been shown that in this context, these three perceptual quality models have the highest correlation against subjective scores, whereas absolute fidelity models like the PSNR correlate poorly with visual perception, producing significantly inferior quality predictions than perception-based quality predictors. It is also worth noting that PSNR was not even considered as an evaluation criteria in the JPEG-AI Call for Evidence \cite{jpegai_data}.

\begin{figure*}[!t]
	\centering
	\footnotesize
	\renewcommand{\tabcolsep}{0pt} 
	\renewcommand{\arraystretch}{1.4} 
	\def\imgwid{0.2502}
	\begin{tabular}{cccc}
    \includegraphics[width=\imgwid\textwidth]{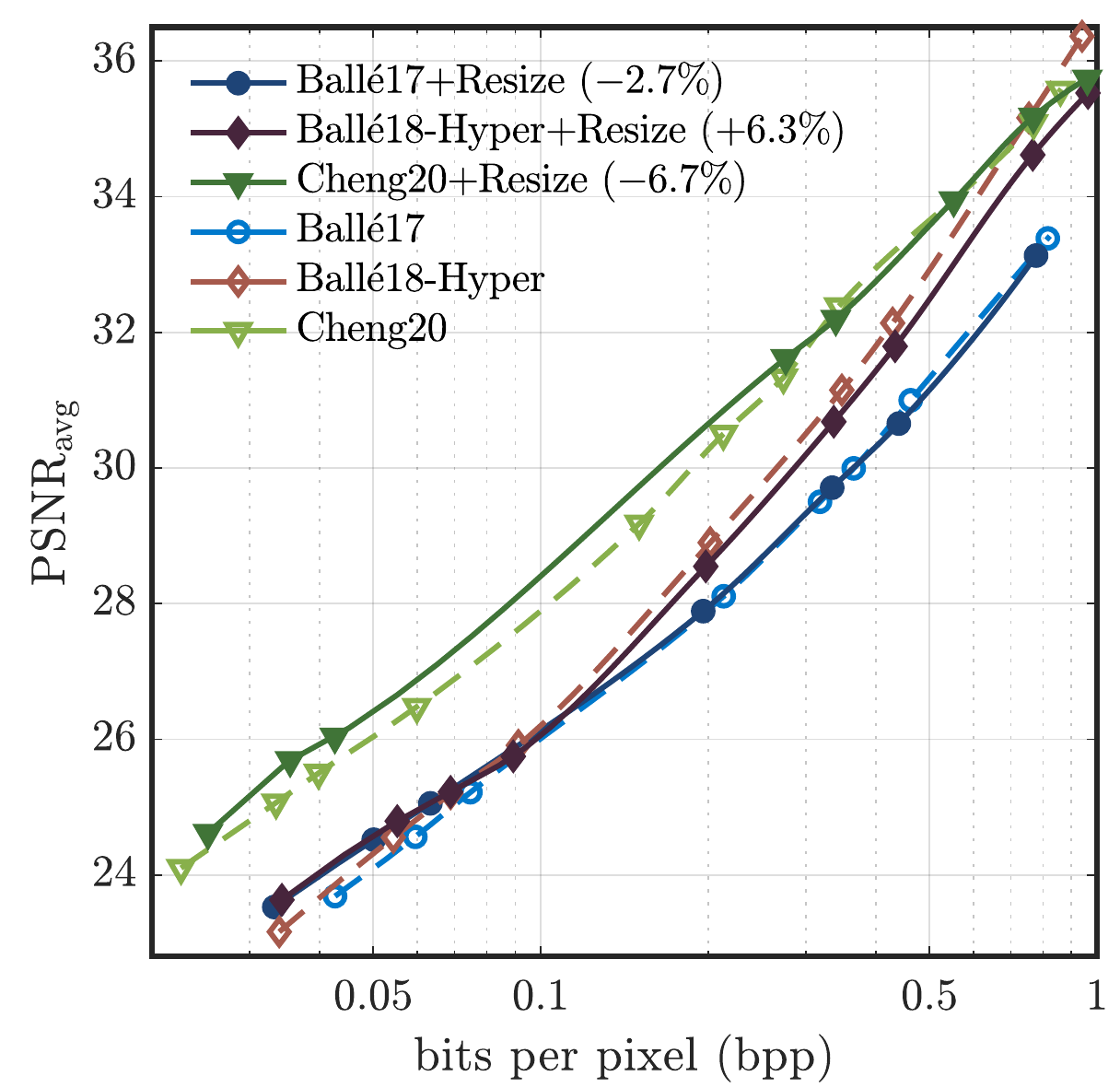} &
    \includegraphics[width=\imgwid\textwidth]{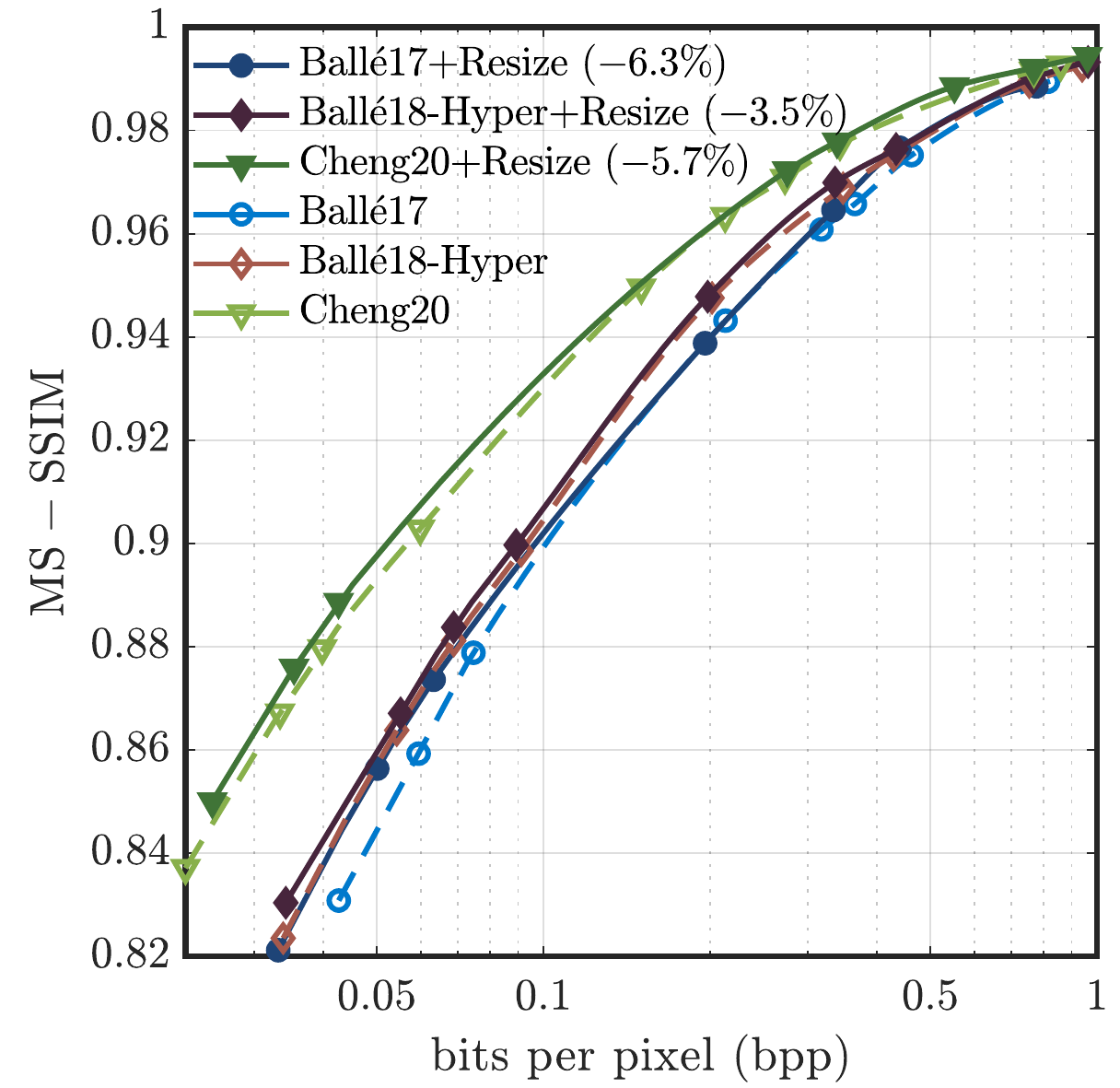} &
    \includegraphics[width=\imgwid\textwidth]{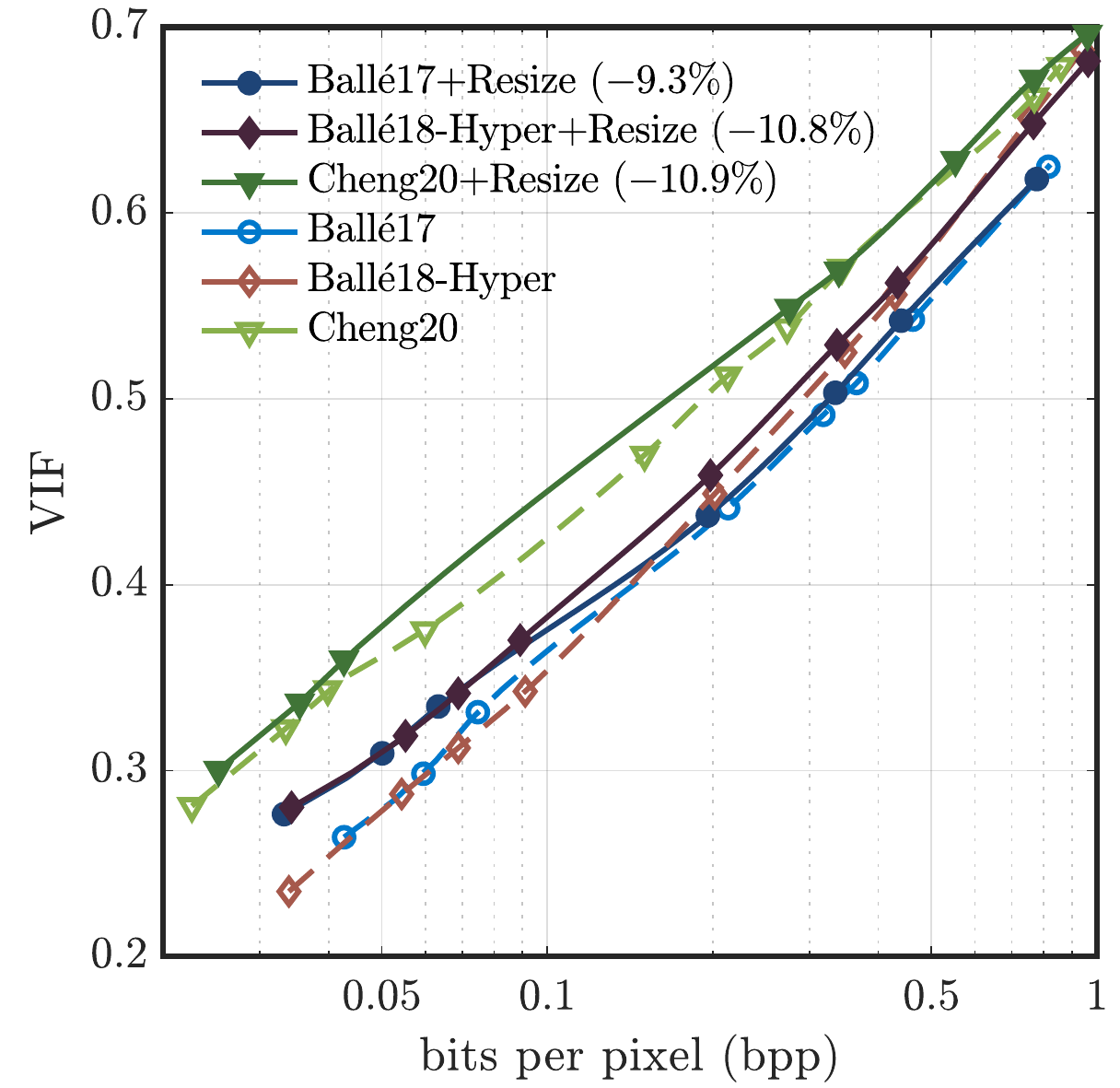} &
    \includegraphics[width=\imgwid\textwidth]{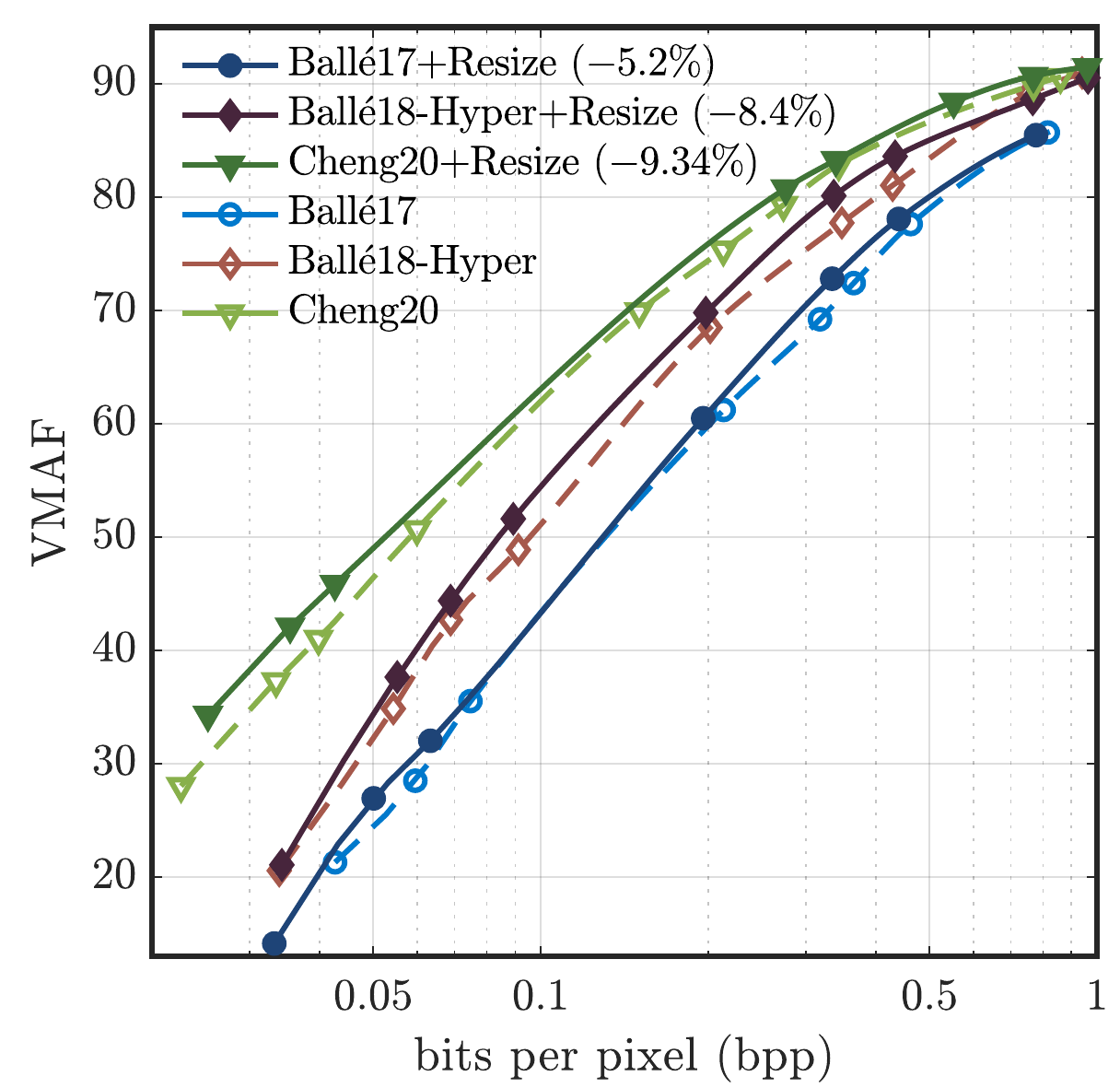} \\
    \multicolumn{4}{c}{(a) \textsc{Kodak} dataset} \\
    \includegraphics[width=\imgwid\textwidth]{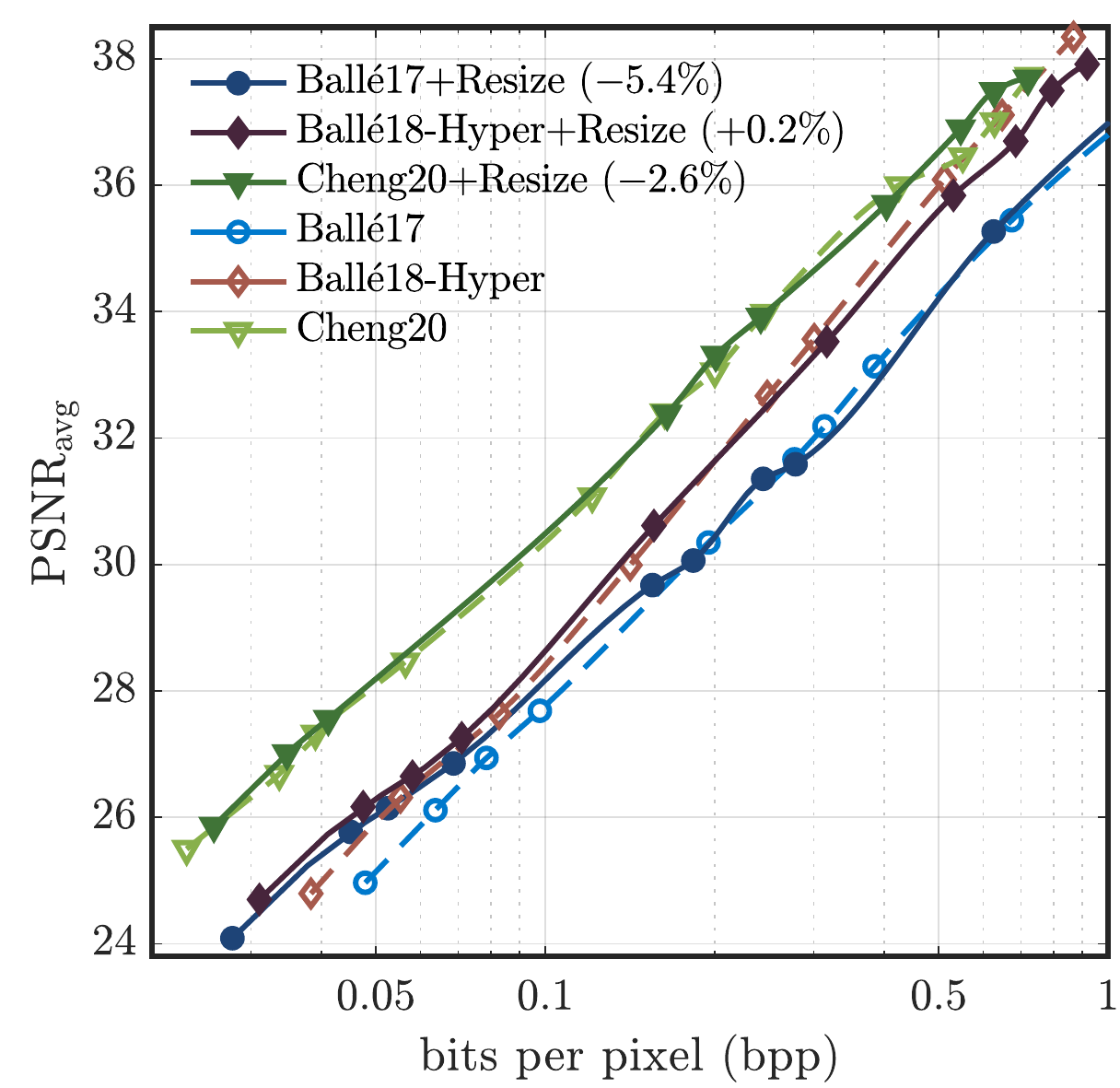} &
    \includegraphics[width=\imgwid\textwidth]{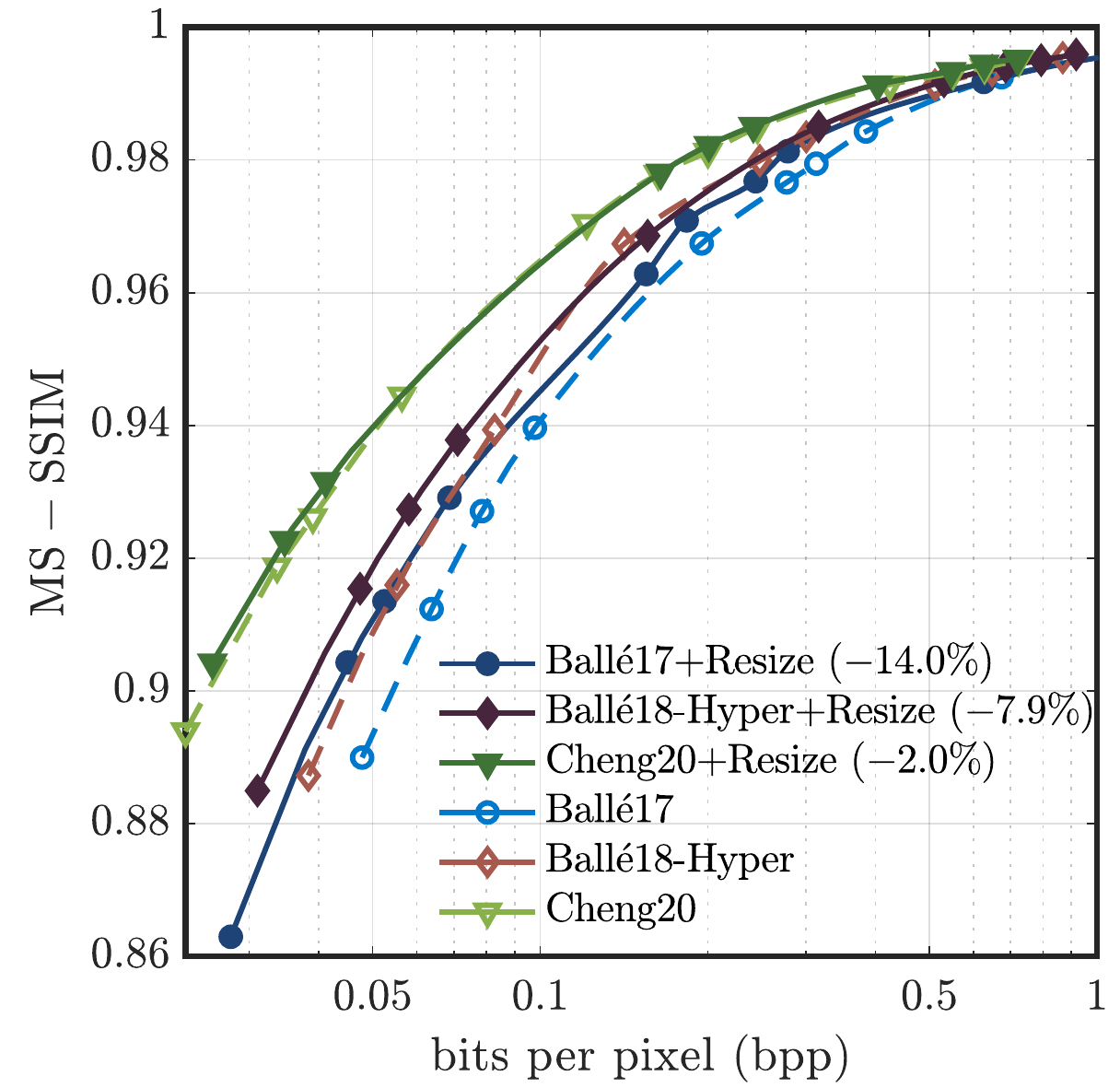} &
    \includegraphics[width=\imgwid\textwidth]{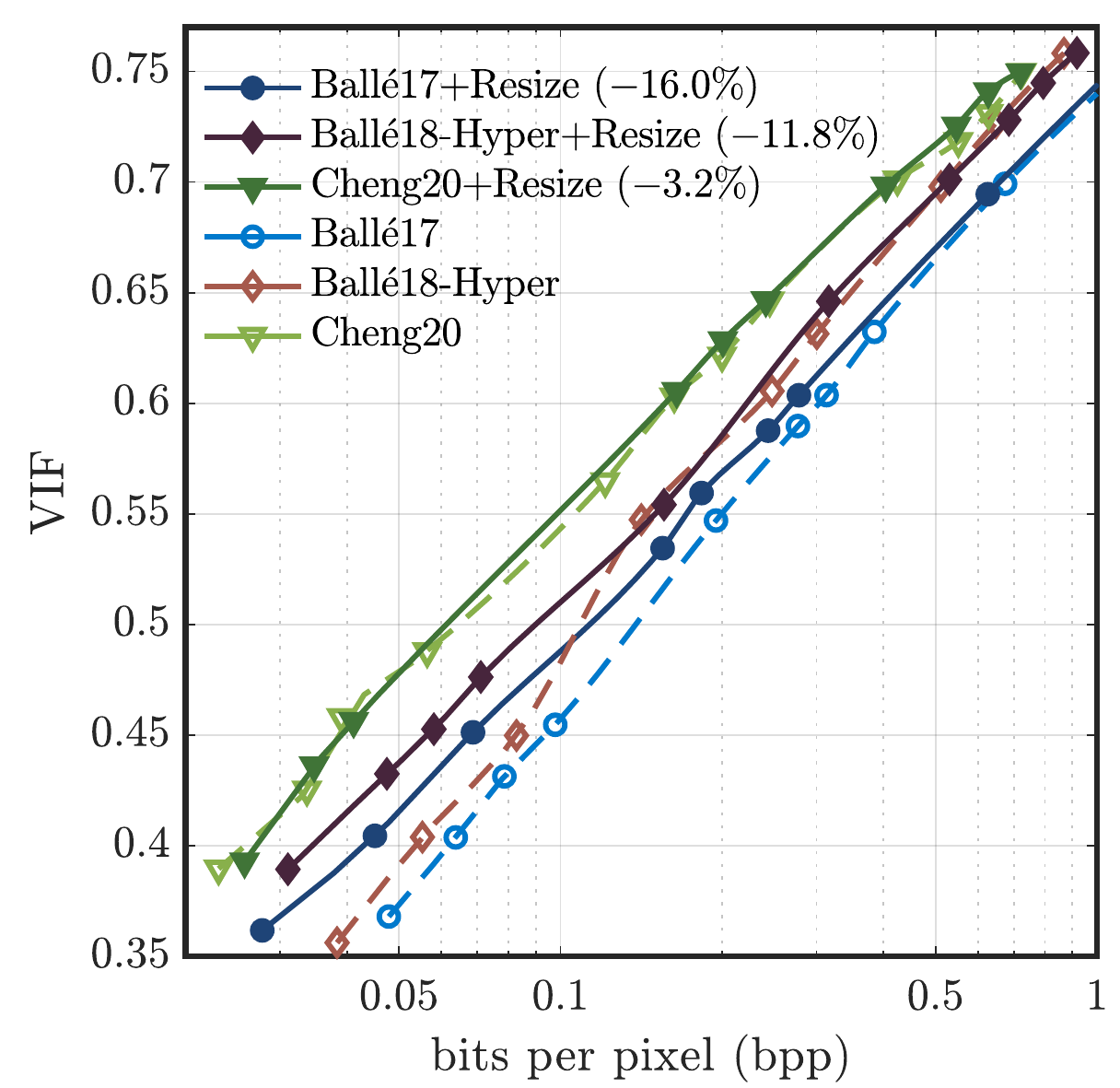} &
    \includegraphics[width=\imgwid\textwidth]{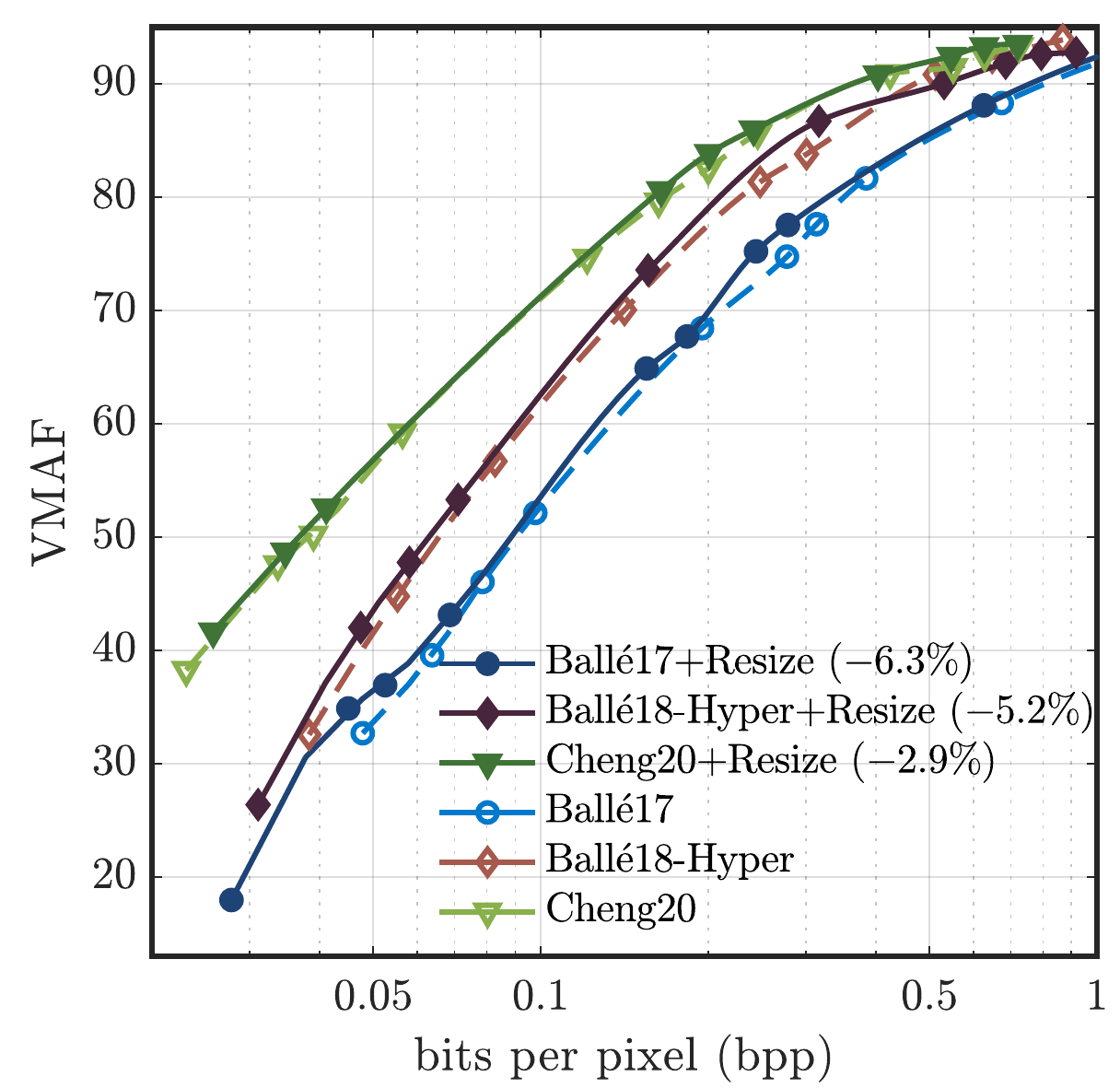} \\
    \multicolumn{4}{c}{(b) \textsc{Tecnick} dataset} \\
    \includegraphics[width=\imgwid\textwidth]{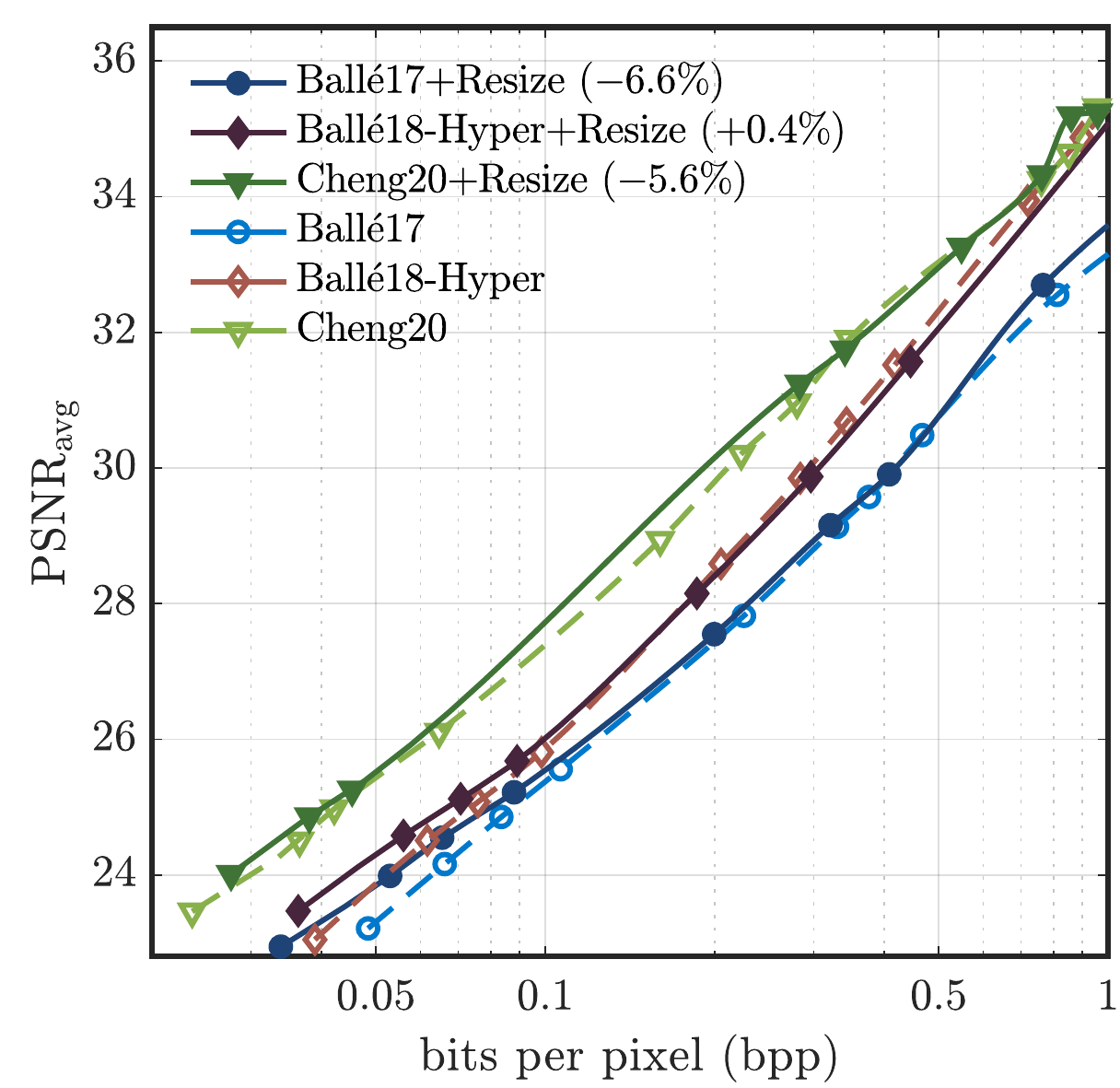} &
    \includegraphics[width=\imgwid\textwidth]{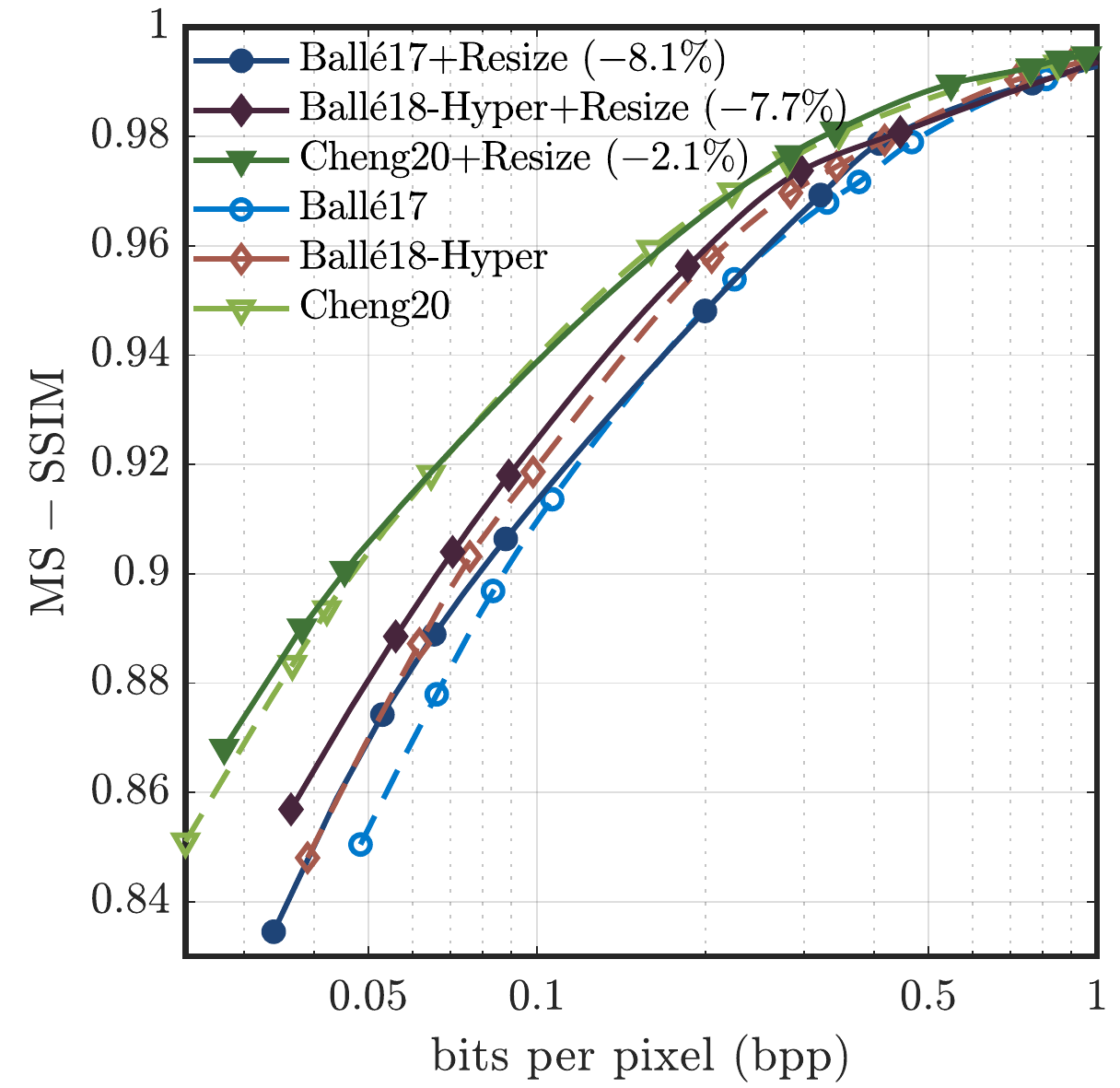} &
    \includegraphics[width=\imgwid\textwidth]{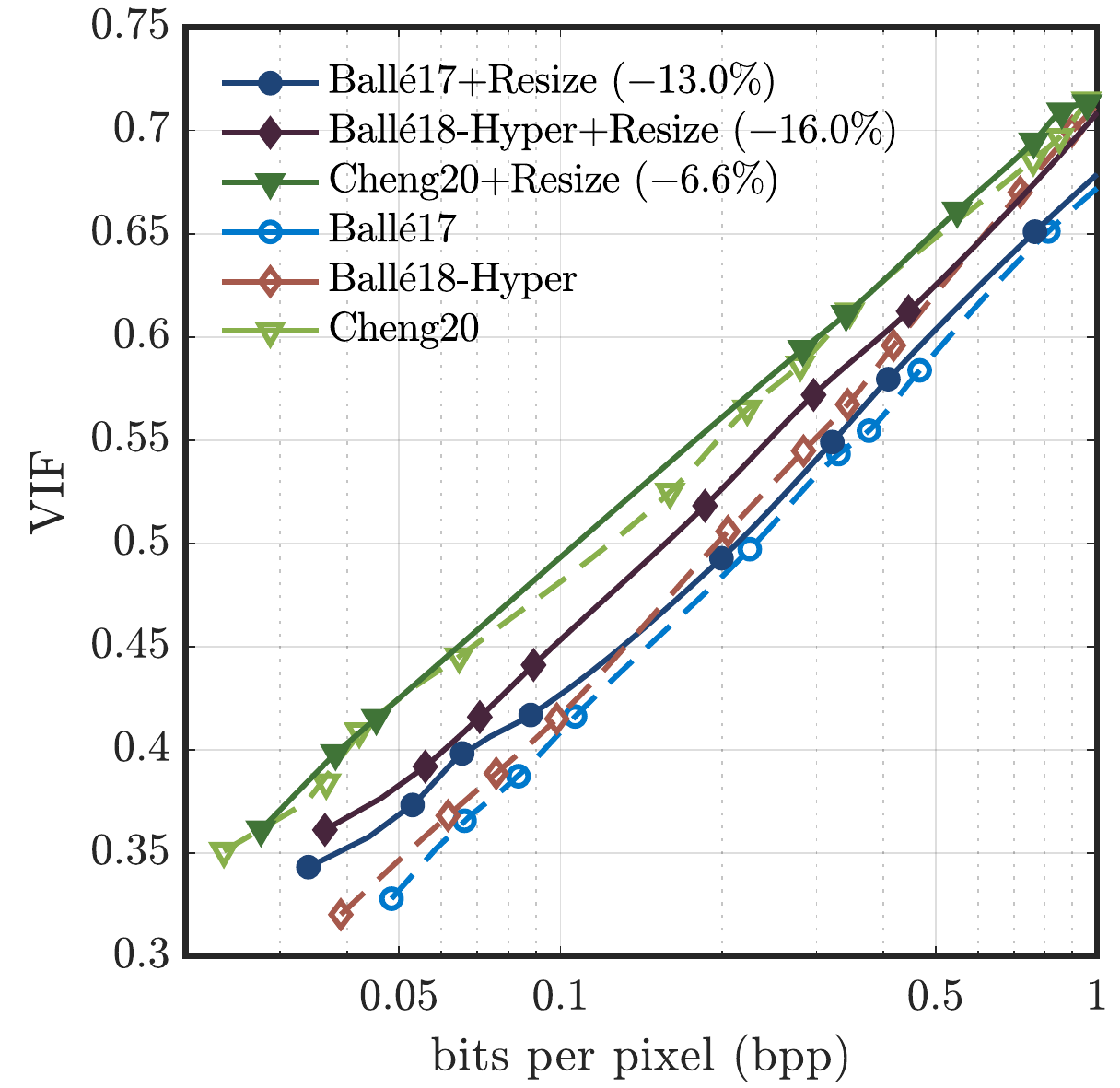} &
    \includegraphics[width=\imgwid\textwidth]{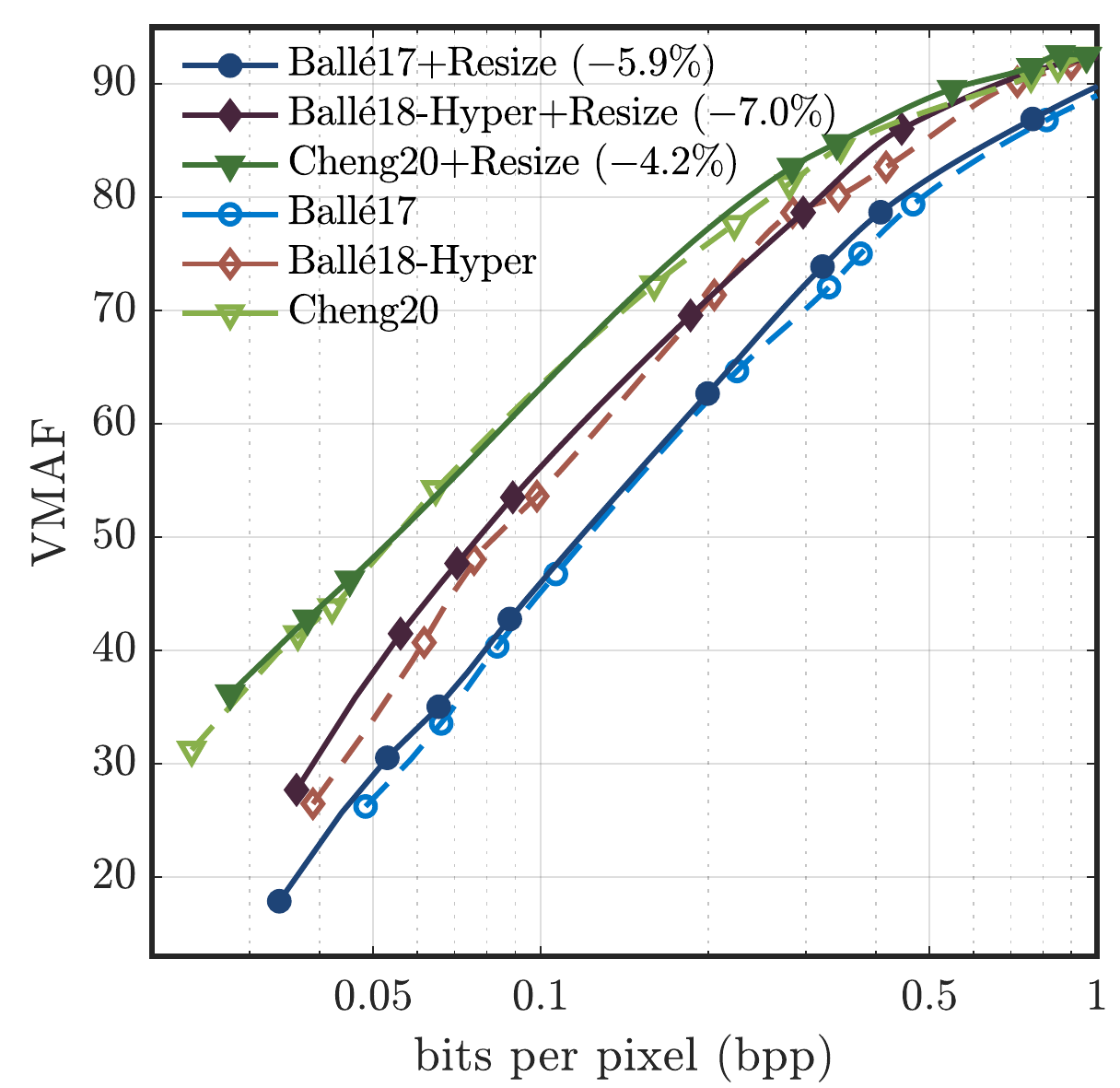} \\
    \multicolumn{4}{c}{(c) \textsc{JPEG AI} dataset} \\
	\end{tabular}
	\caption{Rate-distortion plots aggregated over of the (a) Kodak, (b) {Tecnick}, and (c) {JPEG AI} datasets. The numbers in parentheses are the BD-rates of our \rc~framework against the original compression models. We display the x-axis on a log scale and excluded Ball\'{e}18-Fact for better visualization. The BD-rates shown in the figures were calculated from \textit{C}+Resize against \textit{C}, where \textit{C} represents a compression model.}
	\label{fig:rdcurves}
\end{figure*}

\subsection{Overall Comparison}\label{subs:overall}
We comprehensively evaluated neural compression models against four conventional image codecs: JPEG, JPEG2000, WebP, and the intra-coded HEVC main-RExt (Format Range Extension) profile. For all conventional codecs, we choose maximum gain over encoding speed, and conducted encoding in the codec's native YCbCr color space without chroma subsampling. Extensive experiments were carried out on the three aforementioned datasets, using four representative deep compression models as the backbone to test the generality of our framework: {Ball\'{e}17}\footnotemark{} \cite{BalleLS16a}, {Ball\'{e}18-Fact}\footnotemark[\value{footnote}] (the model using a factorized prior in \cite{balle2018variational}), {Ball\'{e}18-Hyper}\footnotemark[\value{footnote}] (the hyperprior model proposed in \cite{balle2018variational})\footnotetext{https://github.com/tensorflow/compression}, and {Cheng20}\footnote[2]{We implemented the training code based on the network architecture provided by the authors, which has been made available online at https://github.com/treammm/TrainCompression-ChengCVPR2020} \cite{ChengGMM_2020_CVPR}. We tabulated the performances of all of the compared models in Table~\ref{tab:overall}, with respect to different objective image quality assessment models. We report the BD-rate changes obtained relative to the baseline (Ball\'{e}17 \cite{BalleLS16a}), averaged over all the images in the test set. 

As may be observed from the results in the table, our approach (highlighted in \textbf{boldface}) was able to deliver significant BD-rate gains against the original compression model. Interestingly, greater RD performance gains achieved against MS-SSIM, VIF, and VMAF show the perceptual benefits of our proposed \rc~framework to more general perceptual metrics, even though it was trained using MSE loss. For example, despite having similar or marginally worse PSNR BD-rates, we were able to demonstrate that building the early Ball\'{e}17 model on top of our framework (\textbf{Ball\'{e}17+Resize}) achieved similar or better results than the more complex Ball\'{e}18-Fact model, when the results were measured by the perceptual quality models. This suggests that the improvements made were not only in a pixel-wise (PSNR) sense, but that the \resizenet~model also learned to resize images that contributed favorably towards optimizing the visual quality of neural image compression models. It is worth noting that the PSNR BD-rates obtained using our framework performed better particularly on higher resolution test sets (Tecnick and JPEG-AI), possibly because large resolution images are less affected by the boundary cropping described in Section \ref{sec:diff_resize_layer}. In some cases, in particular the Kodak dataset, the BD-rate with respect to the PSNR metric did not always perform as well as the original compression model. This may have been a problem with PSNR, as we will analyze next.

In addition to the overall BD-rate results, we present the aggregated R-D curves on the three test datasets in Fig. \ref{fig:rdcurves} to study coding performance at different bitrates. In this scenario, each R-D point presents the aggregated value calculated by averaging across all test images compressed by the same model with a specific value of  $\lambda$. To be clear, the calculation of R-D performance here is different from that of Table~\ref{tab:overall}, which followed common practice in the MPEG community of averaging over per-image BD-rates. Similar to the trends observed from Table~\ref{tab:overall}, significant coding gains were obtained by applying our \rc~framework, especially as measured by perceptual quality models. As is evident, the benefits obtained by resolution reduction are most significant in low bitrate compression scenarios (e.g., below $0.5$ bpp), while the models with \rc~delivered similar levels of performance at high bitrates. This is not surprising, and has been well demonstrated in early literature \cite{Bruckstein2003,Lin2006,XiaolinWu2009}, since the reduction in quantization error is not large enough to offset the increase of resizing artifacts in high-quality compression. It may also be inferred from the figures that the worse PSNR BD-rate case in the low-resolution Kodak set was caused at high bitrates, likely due to the property of PSNR: when PSNR is large, it is more sensitive to small changes in the MSE. Since our models were trained on high resolution datasets, slight estimation errors of the resize factor may be introduced when testing on low resolution images. Consequently, a small increment in MSE may result in a noticable decrement in PSNR. Yet, this did not affect perceived quality, as attested by the perceptual quality measurements and by the subjective study we will describe later. 

\subsection{Ablation Study}\label{sec:ablation}
In order to study the significance of each module in our framework, we conducted an ablation study by training a series of intermediate models between the original Ball\'{e}17 model \cite{BalleLS16a} and the corresponding using \rc. In this study, all the models were re-trained with 256$\times$256 patch sizes and with the learning rate fixed at $1e^{-4}$ for $1$M steps. For simplicity, we only tested the performance at four different bitrates, ranging from $0.025$ bpp to $0.75$ bpp. We measured changes the R-D performance introduced by removing components from the full model.  

\begin{table}[!t]
  \renewcommand{\arraystretch}{1.4}
  \renewcommand{\tabcolsep}{6.pt} 
  \caption{Ablation experiments on the design of resize layers and resize factor estimation. The baseline comparison is against the Ball\'{e}17+Resize model.}
  \label{tab:ablation}
  \centering
  \begin{tabular}{l c c c c}
  \toprule
  Ablation & PSNR & MS-SSIM & VIF & VMAF\\
  \midrule
  Ball\'{e}17 +Resize \scriptsize{(full model)} 
  & +0.00 & +0.00 & +0.00 & +0.00  \\
  Ball\'{e}17 \cite{BalleLS16a} \scriptsize{(original model)} 
  & +1.48 & +11.29 & +13.93 & +14.00  \\
  \midrule
  (a) w/o bicubic \scriptsize{(w/ bilinear)} & +2.29 & +2.99 & +1.21 & +2.38  \\
  (b) w/o pre-filter & +0.02 & +4.21 & +3.90 & +6.40  \\
  (c) w/o post-filter & +3.87 & +8.31 & +8.14 & +16.70  \\
  (d) = (a) + (b) + (c) & +10.32 & +13.87 & +13.55 & +22.98  \\
  \midrule
  (e) w/o \resizenet & +0.87 & +3.14 & +8.23 & +2.96  \\ 
  (f) est. Mx and My & +12.86 & +2.62 & +2.94 & +5.48  \\
  \bottomrule
  \end{tabular}
\end{table}

\textbf{Study of resize layers} We studied the design choice of resize layers by comparing the four intermediate models below against the full model:
\begin{enumerate}[label=(\alph*)]
  \item Replacing bicubic interpolation in both down-sampling and up-sampling layer with bilinear interpolation;
  \item Removing the pre-filter in the down-sampling layer;
  \item Removing the post-filter in the up-sampling layer;
  \item Removing the pre-filter and the post-filter on top of (a).
\end{enumerate}
From the results in Table \ref{tab:ablation}, we can draw a number of conclusions. The first thing to notice is the drop in performance attained by removing each component. As expected, the best performances were achieved using the full model, indicating that all the modules were important. It may also be observed that adding a post-filter in the up-sampling layer was more effective than using the pre-filter. Among the four cases, the poorest performance in (d) shows that, simply using bilinear interpolation to resize images is insufficient in this context. 

\textbf{Importance of resize parameter estimator} Estimating parameters for image resizing is another essential component of \rc. In this experiment we investigated the significance of the estimation methodology against complexity:
\begin{enumerate}[label=(\alph*)]
  \setcounter{enumi}{4}
  \item Learning a content-agnostic resize factor using a trainable variable $M$ (without \resizenet);
  \item Estimating separate resize parameters $\mathbf{M}=(M_x, M_y)$ for the $x$, $y$ dimensions, respectively. Here $M_x$ and $M_y$ were predicted by two separate CNNs.
\end{enumerate}
In (e), we found that the performance of the content-agnostic model underperformed the full model, since it lacked the flexibility to adjust resize parameter according to the input content. Surprisingly, the result in (f) shows that doubling the number of parameters to conduct independent bidirectional scaling significantly lowered the performance. This may have been due to training instability introduced by an overparameterized model. Thus, we recommend sharing the same resize parameter for both the $x$ and $y$ dimensions.

\begin{table}[!t]
  \renewcommand{\arraystretch}{1.4}
  \renewcommand{\tabcolsep}{3.2pt} 
  \caption{Runtime comparison of deep compression models. All computational speeds are given in milliseconds.}
  \label{tab:exec_time}
  \centering
  \begin{tabular}{l c c c c}
  \toprule
  Compression model & Enc. \scriptsize{(CPU)} & Dec. \scriptsize{(CPU)} & Enc. \scriptsize{(GPU)} & Dec. \scriptsize{(GPU)} \\
  \midrule
  Ball\'{e}17 \cite{BalleLS16a}  
  & 320.62 & 369.65 & 210.16 & 222.73 \\
  Ball\'{e}17~+Resize  
  & 1136.69 & 702.22 & 393.69 & 307.76 \\
  \rowcolor{lightgray!20}
  Ball\'{e}18-Fact \cite{balle2018variational}
  & 824.52 & 1086.26 & 199.46 & 263.77 \\
  \rowcolor{lightgray!20}
  Ball\'{e}18-Fact~+Resize
  & 1686.69 & 1336.37 & 431.51 & 343.75 \\
  Ball\'{e}18-Hyper \cite{balle2018variational}
  & 1006.54 & 1096.46 & 308.52 & 301.09 \\
  Ball\'{e}18-Hyper~+Resize
  & 1781.32 & 1463.94 & 595.75 & 419.84 \\
  \rowcolor{lightgray!20}
  Cheng20 \cite{ChengGMM_2020_CVPR}
  & 46136.25 & 47263.21 & 42506.73 & 44553.21 \\
  \rowcolor{lightgray!20}
  Cheng20~+Resize
  & 46526.51 & 47392.57 & 42736.56 & 45012.58 \\
  \bottomrule
  \end{tabular}
\end{table}

\subsection{Execution Time}
We evaluated the processing times of the various compared deep image compression models in Table \ref{tab:exec_time}. The results were calculated by averaging the runtime (in terms of milliseconds) over all Kodak images on the same machine equipped with an Intel Xeon Platinum 8259CL CPUs@2.50GHz, 128G RAM, and an NVIDIA Tesla K80 GPU. The model loading time was not included. From Table \ref{tab:exec_time}, it may be observed that, due to the integration of \rc, the encoding and decoding complexity was higher as compared to the original models, especially when measured on CPU. However, as seen in the Table, the complexity overhead of our framework was negligible on top of more sophisticated models like Cheng20. Since \resizenet~is not present during decoding, the decoding speed is slightly faster, and we may also infer that the customized differentiable bicubic interpolation is the component that slows the execution speed. This can be further optimized by using native languages such as C++ or by re-compiling with a low-level instruction set.

\subsection{Subjective Study and Analysis}
We conducted an online human subject study to better understand the perceptual preferences of human viewers against different compression models. We selected the Ball\'{e}17 and Ball\'{e}18-Hyper models for the study, and compared them against the corresponding versions equipped with our \rc~framework. We adopted the \textit{two-alternative forced choice (2AFC)} method, since it can be easier for humans to make comparisons between simultaneously displayed pictures having subtle perceptual differences at similar bitrates. We selected all $18$ landscape image contents from the Kodak dataset, where each content was encoded at four different bitrates ranging from $0.01$ bpp to $1$ bpp. To equate bitrates in each comparison, we encoded the images using a 2D 13$\times$13 grid of models at different $\lambda$ values $(\lambda_\text{original}, \lambda_\text{\rc})\in\{\lambda_i\}\times\{\lambda_i\}\vert_{i=1,2,...,13}$, where $\lambda_i\in[0.0001,0.1]$. We selected four $\lambda$ pairs that minimized percent bitrate differences. We customized a user interface to carry out the human study, whereby participants could easily compare pairs of images in their natural environment. On each trial, subjects were shown two images encoded by the original compression model and our \rc~framework, both presented next to the corresponding reference image. After viewing each image triplet, a subject was asked to choose which of the two images had better fidelity compared to the reference. The study included $47$ volunteer participants and about $46\%$ were somewhat knowledgeable about image/video processing, while the others were naive. In sum, we asked each participant to compare $18\times 4\times2=144$ pairs of images, which lasted about $30$-$40$ minutes.

\begin{figure}[!t]
  \centering
  \footnotesize
  \includegraphics[width=3.45in]{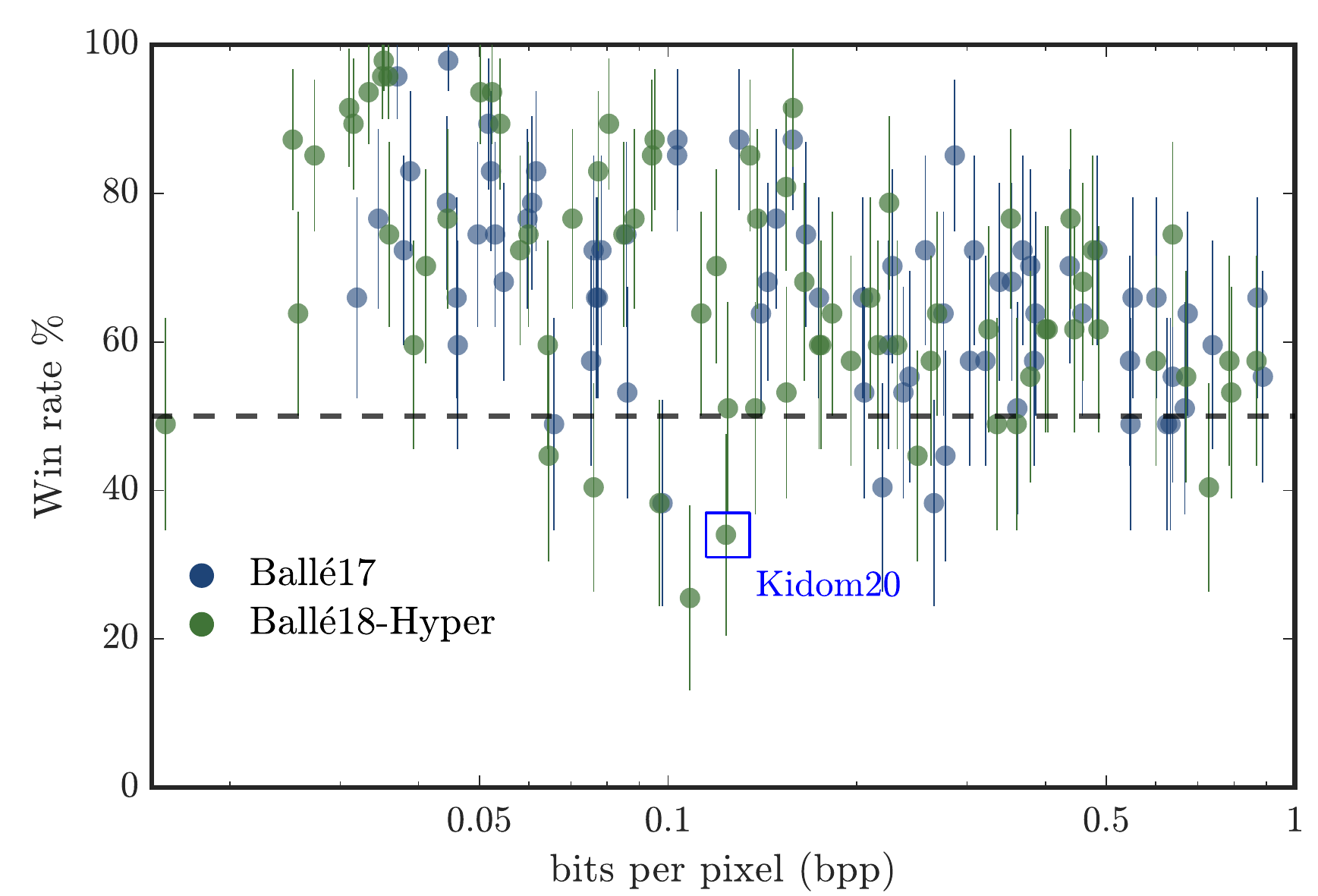}
  \caption{Subjective performance of perceptual win rate of our \rc~framework against the original compression model for each comparison (denoted by $\hat{p}$). The error bars indicate the $95\%$ binomial proportion confidence intervals obtained by $\hat{p}\pm 1.96\sqrt{\hat{p}(1-\hat{p})/n}$, $n=47$.}
  \label{fig:study_winrate}
\end{figure}

Given the collected subjective comparisons, we analyzed the results by computing the percentage of subjects that preferred \rc~over the original model, and plotted the results with respect to the bitrate of each comparison in Fig. \ref{fig:study_winrate}. On average, the images encoded with \rc~framework were preferred by $67.24\%$ of the subjects. Among all the comparisons, $88$ of $144$ cases had win rates significantly larger than $50\%$ while only $2$ cases were worse. It may also be observed that, as bitrate was increased, our approach began to obtain win rates around $50\%$ of the votes, since the distinctions between the codecs then becomes quite subtle.

\begin{figure}[!t]
	\centering
	\footnotesize
	\renewcommand{\tabcolsep}{1.1pt} 
	\renewcommand{\arraystretch}{0.8} 
	\def\imgwid{}
	\begin{tabular}{cc ccc}
    \multicolumn{2}{c}{\rotatebox{90}{~~~~~~~~~~~~~Kodim20 (Source)}} &
    \multicolumn{3}{c}{\includegraphics[width=0.445\textwidth]{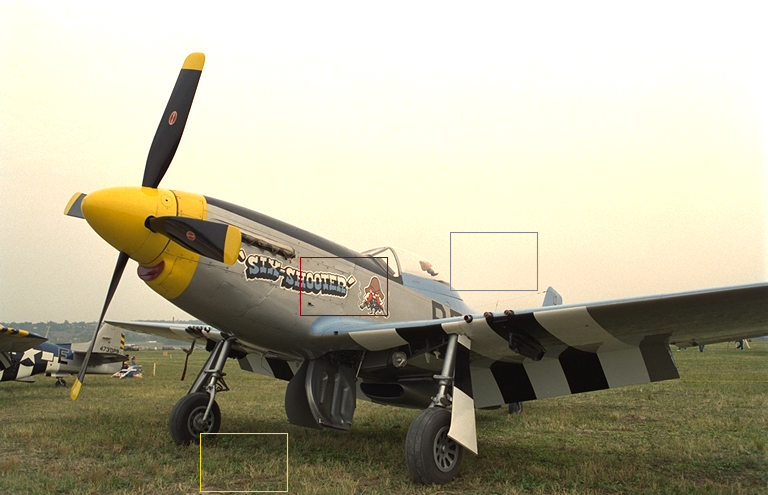}} \\

    \multirow{-3}{*}{\rotatebox{90}{Ball\'{e}18-Hyper}} & 
    \rotatebox{90}{\,~~~original} &

    \colorbox[HTML]{990011}{\makebox(69,45){\includegraphics[width=0.14\textwidth]{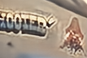}}} &
    \colorbox[HTML]{888789}{\makebox(69,45){\includegraphics[width=0.14\textwidth]{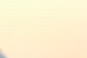}}} &
    \colorbox[HTML]{f5e400}{\makebox(69,45){\includegraphics[width=0.14\textwidth]{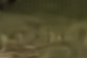}}}\vspace{0.3em}\\
    &
    \rotatebox{90}{~~~~+\textbf{Ours}} &
    \colorbox[HTML]{990011}{\makebox(69,45){\includegraphics[width=0.14\textwidth]{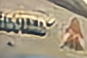}}} &
    \colorbox[HTML]{888789}{\makebox(69,45){\includegraphics[width=0.14\textwidth]{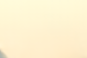}}} &
    \colorbox[HTML]{f5e400}{\makebox(69,45){\includegraphics[width=0.14\textwidth]{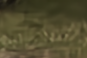}}} \\\\
	\end{tabular}
	\caption{A \textbf{failure case} on image Kodim20: the original Ball\'{e}18-Hyper at 0.135 bpp was preferred by $67\%$ of the subjects over our framework ($33\%$ win rate) at 0.124 bpp.}
	\label{fig:fail_case}
\end{figure}

\textbf{Failure case analysis} While \rc~attained very good performance at low bitrate compression, it did not perform well in a few particular cases. Figure \ref{fig:fail_case} shows a failure case where more subjects rated the result without using our model as better (corresponding to the blue-boxed datapoint in Fig. \ref{fig:study_winrate}). This may have been because the subjects were less forgiving of the blurred logo (highlighted by the \textcolor[HTML]{990011}{red} box) which may have worsened using our approach, and which may have drawn their attention. Conversely, the checkerboard artifacts (\textcolor[HTML]{686769}{gray} box) or blurry (\textcolor[HTML]{c5b400}{yellow} box) created by the original Ball\'{e}18-Hyper model may have been neglected due to their location and relative faintness. These cases further illustrate the challenges of predicting optimal resize factors for compression, given the content diversity of natural images.

\begin{figure}[!t]
  \centering
  \footnotesize
  \includegraphics[width=3.3in]{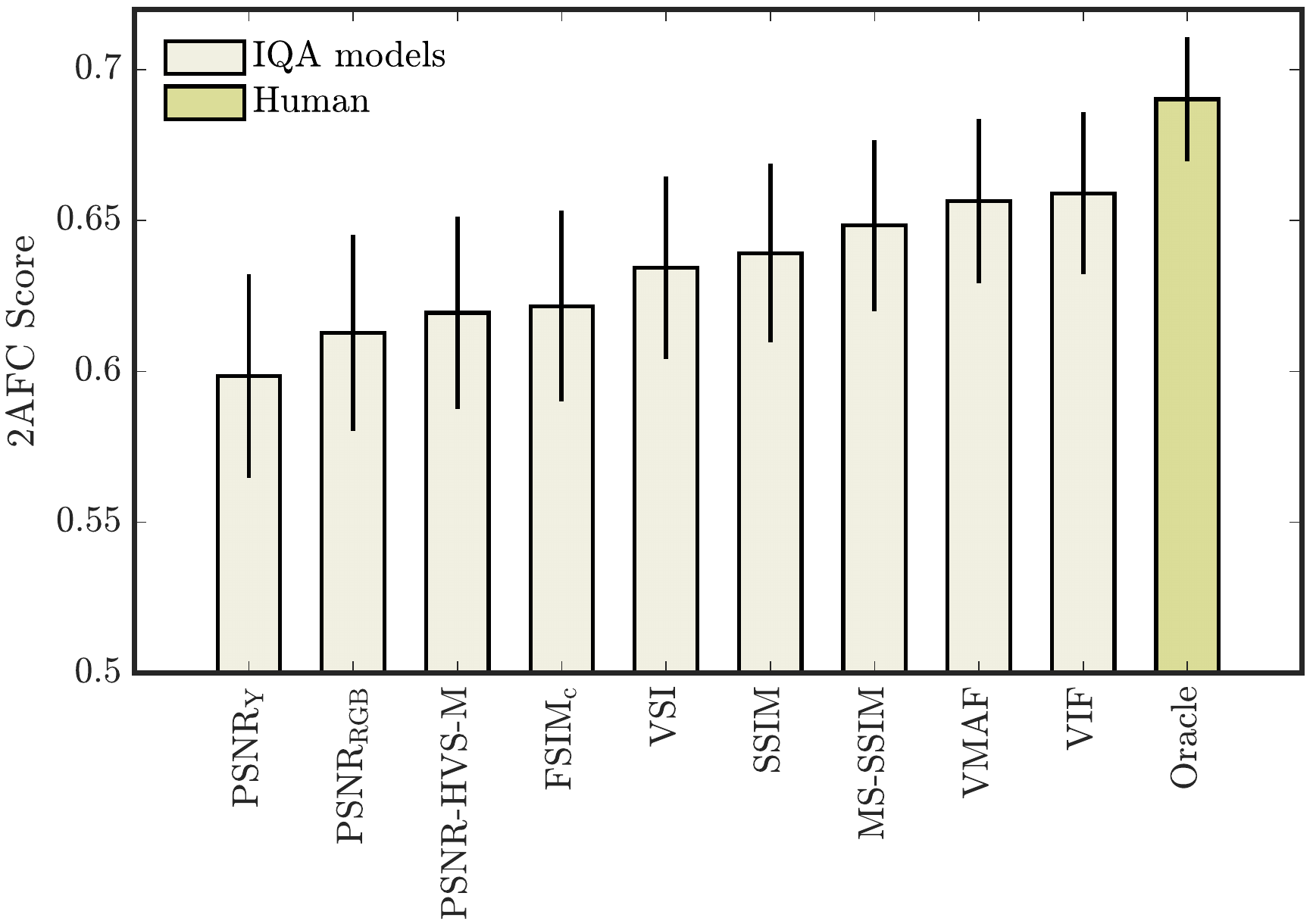}
  \caption{Performance comparison of various IQA methods using the 2AFC score. Larger values indicate better agreement with human judgments.}
  \label{fig:2afc}
\end{figure}

\textbf{2AFC similarity scores} We are also interested in how different image quality assessment (IQA) models correlate with human perception in our study. A standard approach is to compare the correlations between human opinion scores and the predictions made by picture quality models. However, the number of comparisons we obtained in the study is insufficient to estimate the Mean Opinion Score (MOS) using the Bradley-Terry model \cite{btscore1952}. To understand the performance of IQA models under this limitation, we instead adopted the 2AFC scoring method \cite{zhang2018perceptual} given by $\hat{p}\hat{q}+(1-\hat{p})(1-\hat{q})$, where $\hat{p}$ is the percentage of human votes and $\hat{q}\in\{0,1\}$ is the decision made by an IQA model. The idea behind this index is simple: when $\hat{p}$ agrees with $\hat{q}$, the 2AFC score is larger, indicating better performance of a quality model. The theoretical upper bound of the 2AFC score is given by $\max\{\hat{p}, 1-\hat{p}\}$, which can be achieved by an oracle agent $\hat{q}=\mathbbm{1}_{\hat{p}>0.5}$. We evaluated several popular image quality models, including SSIM \cite{WangBSS04}, MS-SSIM \cite{WangMSSSIM03}, PSNR-HVS-M \cite{psnrhvsm07}, VIF \cite{SheikhB06}, VMAF \cite{ZliVMAF18}, FSIM \cite{LZhangFSIM2011}, and VSI \cite{ZhangSLVSI14}, and plotted the 2AFC scores in Fig. \ref{fig:2afc}. Overall, VIF and VMAF achieved the closest performance to the human oracle of $0.69$, whereas MS-SSIM ranked a close third. On the other hand, the levels of performance attained by the PSNR family, which are commonly used in compression, were poor. Like other studies in the literature that analysed the perceptual quality of learning based image compression models \cite{JPEGAI19,Testolina2021}, our experiments also confirmed the value and perceptual relevance using models like MS-SSIM, VIF, and VMAF.

\section{Conclusion and Future Work}
We have introduced a framework for collectively optimizing resize parameter prediction and deep image compression. A distinguishing characteristic of our design is that the optimal resize factor for different contents is estimated by an auxiliary network, without the need for pre-assigned labels. We experimentally demonstrated that, at low bitrate compression, our approach achieved significant improvements in terms of rate-distortion performance, relative to the original compression models. The RD performance gain measured with perceptual quality models and a subjective study further establish the efficacy of optimal resizing as a way of improving perceptual RD tradeoffs and subjective quality. The proposed learned resizing framework is simple and generalizes well across compression models while delivering significant performance improvements. 

The concept of content-adaptive resizing may serve as a tractable and useful tool for perceptually improving the network architectures of other image restoration problems. Our results have also shown that the \resizenet~network produces powerful representations for resizing, with perceptual relevance. Looking ahead, it is possible that the deep features extracted from the learned \resizenet~network could be transferred as a ``fast estimator'' of optimal resize factors in the context of non-differentiable hybrid video codecs.

\section*{Acknowledgment}
The authors would like to thank all the volunteers who took part in the subjective study \cite{TuMAXIM22}. The human study was conducted under the approval of the Institutional Review Board (IRB) under protocol 2007-11-0066.

\ifCLASSOPTIONcaptionsoff
  \newpage
\fi


\bibliographystyle{IEEEtran}
\bibliography{bare_jrnl_TIP_submit}{}

\begin{thebibliography}{10}
\providecommand{\url}[1]{#1}
\csname url@samestyle\endcsname
\providecommand{\newblock}{\relax}
\providecommand{\bibinfo}[2]{#2}
\providecommand{\BIBentrySTDinterwordspacing}{\spaceskip=0pt\relax}
\providecommand{\BIBentryALTinterwordstretchfactor}{4}
\providecommand{\BIBentryALTinterwordspacing}{\spaceskip=\fontdimen2\font plus
\BIBentryALTinterwordstretchfactor\fontdimen3\font minus
  \fontdimen4\font\relax}
\providecommand{\BIBforeignlanguage}[2]{{%
\expandafter\ifx\csname l@#1\endcsname\relax
\typeout{** WARNING: IEEEtran.bst: No hyphenation pattern has been}%
\typeout{** loaded for the language `#1'. Using the pattern for}%
\typeout{** the default language instead.}%
\else
\language=\csname l@#1\endcsname
\fi
#2}}
\providecommand{\BIBdecl}{\relax}
\BIBdecl

\bibitem{winkler2001book}
S.~Winkler, M.~Kunt, and C.~J. van~den Branden~Lambrecht, ``Vision and video:
  Models and applications,'' \emph{Vision Models Appl. Image Video Process.},
  pp. 201--229, 2001.

\bibitem{Uhrina2017}
M.~Uhrina, J.~Bienik, and T.~Mizdos, ``Chroma subsampling influence on the
  perceived video quality for compressed sequences in high resolutions,''
  \emph{Adv. Electr. Electron. Eng.}, vol.~15, no.~4, Nov. 2017.

\bibitem{Mullen85}
K.~T. Mullen, ``The contrast sensitivity of human colour vision to red-green
  and blue-yellow chromatic gratings.'' \emph{J. Physiol.}, vol. 359, pp.
  381--400, 1985.

\bibitem{cvx_pertitle15}
\BIBentryALTinterwordspacing
A.~Aaron, Z.~Li, M.~Manohara, J.~D. Cock, and D.~Ronca, ``Per-title encode
  optimization,'' \emph{The NETFLIX Tech Blog}, 2015. [Online]. Available:
  \url{https://netflixtechblog.com/per-title-encode-optimization-7e99442b62a2}
\BIBentrySTDinterwordspacing

\bibitem{CChen2018}
C.~Chen, Y.-C. Lin, S.~Benting, and A.~Kokaram, ``Optimized transcoding for
  large scale adaptive streaming using playback statistics,'' in \emph{Proc.
  {IEEE} Int. Conf. Image Process.}, Oct. 2018, pp. 3269--3273.

\bibitem{Wu2020}
P.-H. Wu, V.~Kondratenko, and I.~Katsavounidis, ``Fast encoding parameter
  selection for convex hull video encoding,'' in \emph{Proc. SPIE Applications
  Digital Image Process. {XLIII}}, Aug. 2020.

\bibitem{ChenAV12020}
Y.~Chen, D.~Mukherjee, J.~Han, A.~Grange, Y.~Xu, S.~Parker, C.~Chen, H.~Su,
  U.~Joshi, C.-H. Chiang, Y.~Wang, P.~Wilkins, J.~Bankoski, L.~Trudeau,
  N.~Egge, J.-M. Valin, T.~Davies, S.~Midtskogen, A.~Norkin, P.~de~Rivaz, and
  Z.~Liu, ``An overview of coding tools in {AV}1: the first video codec from
  the alliance for open media,'' \emph{{APSIPA} Trans. Signal Inf. Process.},
  vol.~9, 2020.

\bibitem{BrossVVC2021}
B.~Bross, Y.-K. Wang, Y.~Ye, S.~Liu, J.~Chen, G.~J. Sullivan, and J.-R. Ohm,
  ``Overview of the versatile video coding ({VVC}) standard and its
  applications,'' \emph{{IEEE} Trans. Circuits Syst. Video Technol.}, vol.~31,
  no.~10, pp. 3736--3764, Oct. 2021.

\bibitem{JoshiAV1SR2019}
U.~Joshi, D.~Mukherjee, Y.~Chen, S.~Parker, and A.~Grange, ``In-loop frame
  super-resolution in {AV}1,'' in \emph{Proc. IEEE Picture Coding Symp.}, Nov.
  2019.

\bibitem{JVETM0135}
Hendry, Y.-K. Wang, J.~Chen, T.~Davies, A.~Fuldseth, Y.-C. Sun, T.-S. Chang,
  and J.~Lou, \emph{On adaptive resolution change (ARC) for VVC}, Document
  JVET-M0135, ITU-T/ISO/IEC Joint Video Experts Team (JVET), Jan. 2019.

\bibitem{BurgerSH12}
H.~C. Burger, C.~J. Schuler, and S.~Harmeling, ``Image denoising: Can plain
  neural networks compete with {BM3D}?'' in \emph{Proc. IEEE Conf. Comput.
  Vision Pattern Recog.}, Jun. 2012, pp. 2392--2399.

\bibitem{liu2019cyclicgen}
Y.-L. Liu, Y.-T. Liao, Y.-Y. Lin, and Y.-Y. Chuang, ``Deep video frame
  interpolation using cyclic frame generation,'' in \emph{Proc. AAAI}, 2019,
  pp. 8794--8802.

\bibitem{SPaul2020}
S.~Paul, A.~Norkin, and A.~C. Bovik, ``Speeding up {VP}9 intra encoder with
  hierarchical deep learning-based partition prediction,'' \emph{{IEEE} Trans.
  Image Processing}, vol.~29, pp. 8134--8148, 2020.

\bibitem{SPaul2021}
S.~Paul, A.~Norkin, and A.~C. Bovik, ``Self-supervised learning of perceptually
  optimized block motion estimates for video compression,'' \emph{ArXiv}, vol.
  abs/2110.01805, 2021.

\bibitem{BalleLS16a}
J.~Ball\'{e}, V.~Laparra, and E.~P. Simoncelli, ``End-to-end optimized image
  compression,'' in \emph{Proc. Int. Conf. Learn. Represent.}, 2017, pp. 1--27.

\bibitem{TheisCAE17}
L.~Theis, W.~Shi, A.~Cunningham, and F.~Husz\'{e}r, ``Lossy image compression
  with compressive autoencoders,'' in \emph{Proc. Int. Conf. Learn.
  Represent.}, 2017, pp. 1--19.

\bibitem{Toderici2015VariableRI}
G.~Toderici, S.~M. O'Malley, S.~J. Hwang, D.~Vincent, D.~Minnen, S.~Baluja,
  M.~Covell, and R.~Sukthankar, ``Variable rate image compression with
  recurrent neural networks,'' \emph{CoRR}, vol. abs/1511.06085, 2015.

\bibitem{Toderici2017}
G.~Toderici, D.~Vincent, N.~Johnston, S.~J. Hwang, D.~Minnen, J.~Shor, and
  M.~Covell, ``Full resolution image compression with recurrent neural
  networks,'' in \emph{Proc. IEEE Conf. Comput. Vis. Pattern Recog.}, Jul.
  2017, pp. 5306--5314.

\bibitem{Johnston_2018_CVPR}
N.~Johnston, D.~Vincent, D.~Minnen, M.~Covell, S.~Singh, T.~Chinen,
  S.~Jin~Hwang, J.~Shor, and G.~Toderici, ``Improved lossy image compression
  with priming and spatially adaptive bit rates for recurrent networks,'' in
  \emph{Proc. IEEE Conf. Comput. Vis. Pattern Recog.}, June 2018, pp.
  4385--4393.

\bibitem{agustsson2019generative}
E.~Agustsson, M.~Tschannen, F.~Mentzer, R.~Timofte, and L.~V. Gool,
  ``Generative adversarial networks for extreme learned image compression,'' in
  \emph{Proc. {IEEE} Int. Conf. Comput. Vis.}, 2019, pp. 221--231.

\bibitem{Lhdefink2019GANVJ}
J.~L{\"o}hdefink, A.~B{\"a}r, N.~M. Schmidt, F.~H{\"u}ger, P.~Schlicht, and
  T.~Fingscheidt, ``{GAN}- vs. {JPEG}2000 image compression for distributed
  automotive perception: Higher peak snr does not mean better semantic
  segmentation,'' \emph{ArXiv}, vol. abs/1902.04311, 2019.

\bibitem{mentzer2020high}
F.~Mentzer, G.~D. Toderici, M.~Tschannen, and E.~Agustsson, ``High-fidelity
  generative image compression,'' in \emph{Proc. Adv. Neural Inf. Process.
  Syst.}, vol.~33, 2020, pp. 11\,913--11\,924.

\bibitem{Nakanishi2019}
K.~M. Nakanishi, S.~ichi Maeda, T.~Miyato, and D.~Okanohara, ``Neural
  multi-scale image compression,'' in \emph{Proc. Asia. Conf. Comput. Vis.},
  2019, pp. 718--732.

\bibitem{MaIWave2020}
H.~Ma, D.~Liu, N.~Yan, H.~Li, and F.~Wu, ``End-to-end optimized versatile image
  compression with wavelet-like transform,'' \emph{{IEEE} Trans. Pattern Anal.
  Mach. Intell.}, pp. 1--1, 2020.

\bibitem{balle2018variational}
J.~Ball\'{e}, D.~Minnen, S.~Singh, S.~J. Hwang, and N.~Johnston, ``Variational
  image compression with a scale hyperprior,'' in \emph{Proc. Int. Conf. Learn.
  Represent.}, 2018, pp. 1--23.

\bibitem{NIPS2018_8275}
D.~Minnen, J.~Ball\'{e}, and G.~D. Toderici, ``Joint autoregressive and
  hierarchical priors for learned image compression,'' in \emph{Proc. Adv.
  Neural Inf. Process. Syst.}, 2018, pp. 10\,771--10\,780.

\bibitem{Lee2019Context}
J.~Lee, S.~Cho, and S.-K. Beack, ``Context-adaptive entropy model for
  end-to-end optimized image compression,'' in \emph{Proc. Int. Conf. Learn.
  Represent.}, 2019.

\bibitem{Mentzer_2018_CVPR}
F.~Mentzer, E.~Agustsson, M.~Tschannen, R.~Timofte, and L.~V. Gool,
  ``Conditional probability models for deep image compression,'' in \emph{Proc.
  IEEE Conf. Comput. Vision Pattern Recog.}, Jul. 2018, pp. 4394--4402.

\bibitem{ChenContext2021}
T.~Chen, H.~Liu, Z.~Ma, Q.~Shen, X.~Cao, and Y.~Wang, ``End-to-end learnt image
  compression via non-local attention optimization and improved context
  modeling,'' \emph{{IEEE} Trans. Image Process.}, vol.~30, pp. 3179--3191,
  2021.

\bibitem{ChengGMM_2020_CVPR}
Z.~Cheng, H.~Sun, M.~Takeuchi, and J.~Katto, ``Learned image compression with
  discretized gaussian mixture likelihoods and attention modules,'' in
  \emph{Proc. IEEE Conf. Comput. Vision Pattern Recog.}, Jul. 2020, pp.
  7939--7948.

\bibitem{wu2018vcii}
C.-Y. Wu, N.~Singhal, and P.~Kr{\"a}henb{\"u}hl, ``Video compression through
  image interpolation,'' in \emph{Proc. Eur. Conf. Comput. Vis.}, 2018.

\bibitem{cheng19}
Z.~Cheng, H.~Sun, M.~Takeuchi, and J.~Katto, ``Learning image and video
  compression through spatial-temporal energy compaction,'' in \emph{Proc. IEEE
  Conf. Comput. Vis. Pattern Recog.}, 2019.

\bibitem{LuDVC2019CVPR}
G.~Lu, W.~Ouyang, D.~Xu, X.~Zhang, C.~Cai, and Z.~Gao, ``{DVC}: An end-to-end
  deep video compression framework,'' in \emph{Proc. IEEE Conf. Comput. Vision
  Pattern Recog.}, Jul. 2019, pp. 11\,006--11\,015.

\bibitem{rippel2019iccv}
O.~Rippel, S.~Nair, C.~Lew, S.~Branson, A.~G. Anderson, and L.~Bourdev,
  ``Learned video compression,'' in \emph{Proc. {IEEE} Int. Conf. Comput.
  Vis.}, Sep. 2019, pp. 3454--3463.

\bibitem{djelouah12019iccv}
A.~Djelouah, J.~Campos, S.~Schaub-Meyer, and C.~Schroers, ``Neural inter-frame
  compression for video coding,'' in \emph{Proc. {IEEE} Int. Conf. Comput.
  Vis.}, Sep. 2019, pp. 6421--6429.

\bibitem{agustsson2020CVPR}
E.~Agustsson, D.~Minnen, N.~Johnston, J.~Ball{\'{e}}, S.~J. Hwang, and
  G.~Toderici, ``Scale-space flow for end-to-end optimized video compression,''
  in \emph{Proc. IEEE Conf. Comput. Vision Pattern Recog.}, Jul. 2020, pp.
  8503--8512.

\bibitem{liuNVC2021}
H.~Liu, M.~Lu, Z.~Ma, F.~Wang, Z.~Xie, X.~Cao, and Y.~Wang, ``Neural video
  coding using multiscale motion compensation and spatiotemporal context
  model,'' \emph{arXiv preprint arXiv:2007.04574}, 2020.

\bibitem{MXChenPCS2021}
M.~Chen, A.~Patney, and A.~C. Bovik, ``{MOVI}-codec: Deep video compression
  without motion,'' in \emph{Proc. IEEE Picture Coding Symp.}, Jun. 2021, pp.
  51--55.

\bibitem{NIPS2015_33ceb07b}
M.~Jaderberg, K.~Simonyan, A.~Zisserman, and K.~Kavukcuoglu, ``Spatial
  transformer networks,'' in \emph{Proc. Adv. Neural Inf. Process. Syst.},
  2015, pp. 2017--2025.

\bibitem{Bruckstein2003}
A.~Bruckstein, M.~Elad, and R.~Kimmel, ``Down-scaling for better transform
  compression,'' \emph{{IEEE} Trans. Image Process.}, vol.~12, no.~9, pp.
  1132--1144, Sep. 2003.

\bibitem{Lin2006}
W.~Lin and L.~Dong, ``Adaptive downsampling to improve image compression at low
  bit rates,'' \emph{{IEEE} Trans. Image Process.}, vol.~15, no.~9, pp.
  2513--2521, Sep. 2006.

\bibitem{XiaolinWu2009}
X.~Wu, X.~Zhang, and X.~Wang, ``Low bit-rate image compression via adaptive
  down-sampling and constrained least squares upconversion,'' \emph{{IEEE}
  Trans. Image Process.}, vol.~18, no.~3, pp. 552--561, Mar. 2009.

\bibitem{Knoche2005}
M.~Bhat, J.-M. Thiesse, and P.~L. Callet, ``Can small be beautiful?: Assessing
  image resolution requirements for mobile tv,'' in \emph{Proc. 13th Annu. ACM
  Int. Conf. Multimedia}, Nov. 2005, pp. 829--838.

\bibitem{Cermak2011}
G.~Cermak, M.~Pinson, and S.~Wolf, ``The relationship among video quality,
  screen resolution, and bit rate,'' \emph{{IEEE} Trans Broadcast.}, vol.~57,
  no.~2, pp. 258--262, Jun. 2011.

\bibitem{Georgis2016}
G.~Georgis, G.~Lentaris, and D.~Reisis, ``Reduced complexity superresolution
  for low-bitrate video compression,'' \emph{{IEEE} Trans. Circuits Syst. Video
  Technol.}, vol.~26, no.~2, pp. 332--345, Feb. 2016.

\bibitem{Toni2015}
L.~Toni, R.~Aparicio-Pardo, K.~Pires, G.~Simon, A.~Blanc, and P.~Frossard,
  ``Optimal selection of adaptive streaming representations,'' \emph{{ACM}
  Trans. Multimedia Comput. Commun. Appl.}, vol.~11, no.~2s, pp. 1--26, Feb.
  2015.

\bibitem{Li2016}
C.~Li, L.~Toni, P.~Frossard, H.~Xiong, and J.~Zou, ``Complexity constrained
  representation selection for dynamic adaptive streaming,'' in \emph{Proc.
  {IEEE} Vis. Commun. Image Process.}, Nov. 2016.

\bibitem{Sani2017}
Y.~Sani, A.~Mauthe, and C.~Edwards, ``Adaptive bitrate selection: A survey,''
  \emph{{IEEE} Commun. Surv. Tutor}, vol.~19, no.~4, pp. 2985--3014, 2017.

\bibitem{Afonso2019}
M.~Afonso, F.~Zhang, and D.~R. Bull, ``Video compression based on
  spatio-temporal resolution adaptation,'' \emph{{IEEE} Trans. Circuits Syst.
  Video Technol.}, vol.~29, no.~1, pp. 275--280, Jan. 2019.

\bibitem{Zhang19ViSTRA2}
F.~Zhang, M.~Afonso, and D.~R. Bull, ``{ViSTRA}2: Video coding using spatial
  resolution and effective bit depth adaptation,'' \emph{Signal Process. Image
  Commun.}, vol.~97, p. 116355, Sep. 2021.

\bibitem{Bhat2020}
M.~Bhat, J.-M. Thiesse, and P.~L. Callet, ``A case study of machine learning
  classifiers for real-time adaptive resolution prediction in video coding,''
  in \emph{Proc. {IEEE} Int. Conf. Multimedia Expo}, Jul. 2020, pp. 1--6.

\bibitem{MinminShen2011}
M.~Shen, P.~Xue, and C.~Wang, ``Down-sampling based video coding using
  super-resolution technique,'' \emph{{IEEE} Trans. Circuits Syst. Video
  Technol.}, vol.~21, no.~6, pp. 755--765, Jun. 2011.

\bibitem{HanAV12021}
J.~Han, B.~Li, D.~Mukherjee, C.-H. Chiang, A.~Grange, C.~Chen, H.~Su,
  S.~Parker, S.~Deng, U.~Joshi, Y.~Chen, Y.~Wang, P.~Wilkins, Y.~Xu, and
  J.~Bankoski, ``A technical overview of {AV}1,'' \emph{Proceedings of the
  {IEEE}}, vol. 109, no.~9, pp. 1435--1462, Sep. 2021.

\bibitem{ChangARC2020}
T.-S. Chang, Y.-C. Sun, L.~Zhu, and J.~Lou, ``Adaptive resolution change for
  versatile video coding,'' in \emph{Proc. {IEEE} Vis. Commun. Image Process.},
  Dec. 2020.

\bibitem{FuRPR2021}
T.~Fu, K.~Zhang, L.~Zhang, S.~Wang, and S.~Ma, ``An enhanced reference
  structure for reference picture resampling ({RPR}) in {VVC},'' in \emph{Proc.
  {IEEE} Int. Conf. Image Process.}, Sep. 2021.

\bibitem{Albreem2021}
M.~A. Albreem, A.~H.~A. Habbash, A.~M. Abu-Hudrouss, and S.~S. Ikki, ``Overview
  of precoding techniques for massive {MIMO},'' \emph{{IEEE} Access}, vol.~9,
  pp. 60\,764--60\,801, 2021.

\bibitem{Dong2016}
C.~Dong, C.~C. Loy, K.~He, and X.~Tang, ``Image super-resolution using deep
  convolutional networks,'' \emph{{IEEE} Trans. Pattern Anal. Mach. Intell.},
  vol.~38, no.~2, pp. 295--307, Feb. 2016.

\bibitem{HeZRS16}
K.~He, X.~Zhang, S.~Ren, and J.~Sun, ``Deep residual learning for image
  recognition,'' in \emph{Proc. IEEE Conf. Comput. Vis. Pattern Recog.}, Jun.
  2016, pp. 770--778.

\bibitem{kodak_data}
\BIBentryALTinterwordspacing
\emph{Kodak lossless true color image suite}, 2007. [Online]. Available:
  \url{https://r0k.us/graphics/kodak/}
\BIBentrySTDinterwordspacing

\bibitem{stag_tecnick}
N.~Asuni and A.~Giachetti, ``{TESTIMAGES}: a large-scale archive for testing
  visual devices and basic image processing algorithms,'' in \emph{Proc.
  Eurographics Italian Chapter Conference}, 2014, pp. 63--70.

\bibitem{jpegai_data}
\BIBentryALTinterwordspacing
\emph{JPEG-AI Call for Evidence - IEEE MMSP2020 challenge (dataset)}, 2020.
  [Online]. Available: \url{http://jpegai.github.io/test\_images/}
\BIBentrySTDinterwordspacing

\bibitem{clic_data}
\BIBentryALTinterwordspacing
\emph{Dataset of the CVPR Workshop and Challenge on Learned Image Compression
  (CLIC)}, 2020. [Online]. Available: \url{http://clic.compression.cc/}
\BIBentrySTDinterwordspacing

\bibitem{kingma:adam}
D.~P. Kingma and J.~Ba, ``Adam: A method for stochastic optimization,'' in
  \emph{Proc. Int. Conf. Learn. Represent.}, 2015, pp. 1--15.

\bibitem{BDRate01}
{G. Bj\o{}ntegaard}, ``Calculation of average {PSNR} differences between
  {RD}-curves,'' \emph{Document VCEG-M33, ITU-T Video Coding Experts Group
  (VCEG) Thirteenth Meeting}, Austin, TX, April 2001.

\bibitem{WangMSSSIM03}
Z.~Wang, E.~P. Simoncelli, and A.~C. Bovik, ``Multi-scale structural similarity
  for image quality assessment,'' in \emph{Proc. IEEE Asilomar Conf. on
  Signals, Syst., and Comput.}, Nov. 2003, pp. 1398--1402.

\bibitem{SheikhB06}
H.~R. Sheikh and A.~C. Bovik, ``Image information and visual quality,''
  \emph{IEEE Trans. Image Process.}, vol.~15, no.~2, pp. 430--444, Feb. 2006.

\bibitem{ZliVMAF18}
\BIBentryALTinterwordspacing
Z.~Li, C.~Bampis, J.~Novak, A.~Aaron, K.~Swanson, A.~Moorthy, and J.~D. Cock,
  ``{VMAF}: The journey continues,'' \emph{The NETFLIX Tech Blog}, 2018.
  [Online]. Available:
  \url{https://netflixtechblog.com/vmaf-the-journey-continues-44b51ee9ed12}
\BIBentrySTDinterwordspacing

\bibitem{JPEGAI19}
J.~Ascenso, P.~Akayzi, M.~Testolina, A.~Boev, and E.~Alshina, \emph{Performance
  evaluation of learning based image coding solutions and quality metrics},
  Document ISO/IEC JTC 1/SC29/WG1 N85013, 85th JPEG Meeting, Nov. 2019.

\bibitem{Testolina2021}
M.~Testolina, E.~Upenik, J.~Ascenso, F.~Pereira, and T.~Ebrahimi, ``Performance
  evaluation of objective image quality metrics on conventional and
  learning-based compression artifacts,'' in \emph{Proc. 13th Int. Conf.
  Quality Multimedia Exper.}, Jun. 2021.

\bibitem{btscore1952}
R.~A. Bradley and M.~E. Terry, ``Rank analysis of incomplete block designs: The
  method of paired comparisons,'' \emph{Biometrika}, vol.~39, no. 3-4, pp.
  324--345, Dec. 1952.

\bibitem{zhang2018perceptual}
R.~Zhang, P.~Isola, A.~A. Efros, E.~Shechtman, and O.~Wang, ``The unreasonable
  effectiveness of deep features as a perceptual metric,'' in \emph{Proc. IEEE
  Conf. Comput. Vision Pattern Recog.}, Jun. 2018, pp. 586--595.

\bibitem{WangBSS04}
Z.~Wang, A.~Bovik, H.~Sheikh, and E.~Simoncelli, ``Image quality assessment:
  From error visibility to structural similarity,'' \emph{IEEE Trans. Image
  Process.}, vol.~13, no.~4, pp. 600--612, Apr. 2004.

\bibitem{psnrhvsm07}
N.~Ponomarenko, F.~Silvestri, K.~Egiazarian, M.~Carli, J.~Astola, and V.~Lukin,
  ``On between-coefficient contrast masking of {DCT} basis functions,'' in
  \emph{3rd Int. Workshop Video Process. Quality Metrics Consum. Electron.},
  Jan. 2007.

\bibitem{LZhangFSIM2011}
L.~Zhang, L.~Zhang, X.~Mou, and D.~Zhang, ``{FSIM}: A feature similarity index
  for image quality assessment,'' \emph{IEEE Trans. Image Process.}, vol.~20,
  no.~8, pp. 2378--2386, Aug. 2011.

\bibitem{ZhangSLVSI14}
L.~Zhang, Y.~Shen, and H.~Li, ``{VSI}: A visual saliency-induced index for
  perceptual image quality assessment,'' \emph{IEEE Trans. Image Process.},
  vol.~23, no.~10, pp. 4270--4281, Oct. 2014.

\bibitem{TuMAXIM22}
Z.~Tu, H.~Talebi, H.~Zhang, F.~Yang, P.~Milanfar, A.~Bovik, and Y.~Li, ``Maxim:
  Multi-axis mlp for image processing,'' \emph{ArXiv}, vol. abs/2201.02973,
  2022.

\end{thebibliography}

\end{document}